\documentclass[oneside,british,latin9,usenames,table,dvipsnames,svgnames]{aa}
\usepackage[T1]{fontenc}
\usepackage{geometry}
\usepackage{color}
\usepackage{babel}
\usepackage{float}
\usepackage{textcomp}
\usepackage{url}
\usepackage{amstext}
\usepackage{amssymb}
\usepackage{graphicx}
\usepackage[unicode=true,pdfusetitle,
 bookmarks=true,bookmarksnumbered=false,bookmarksopen=false,
 breaklinks=false,pdfborder={0 0 0},pdfborderstyle={},backref=false,colorlinks=true]
 {hyperref}
\hypersetup{
 urlcolor=blue,citecolor=blue}

\makeatletter

\providecommand{\tabularnewline}{\\}

\floatstyle{ruled}
\newfloat{algorithm}{tbp}{loa}
\providecommand{\algorithmname}{Algorithm}
\floatname{algorithm}{\protect\algorithmname}

\newcommand\mathrange{\textrm{--}}

\newcommand\XARS{{\texttt{XARS}}}
\newcommand\xspec{{\texttt{Xspec}}}

\newcommand{\NH}{ {N_{\rm{H}}} }
\newcommand{\NHunit}{ \,{\mathrm{cm}^{-2}} }

\newcommand{\chiwarp}{1432}
\newcommand{\chimytorus}{1502}
\newcommand{\chictorus}{1523}
\newcommand{\chiwadaring}{1451}
\newcommand{\chiwada}{4392}
\newcommand{\chidofwarp}{\chiwarp/1462}
\newcommand{\chidofmytorus}{\chimytorus/1463}
\newcommand{\chidofctorus}{\chictorus/1462}
\newcommand{\chidofwadaring}{\chiwadaring/1463}
\newcommand{\chidofwada}{\chiwada/1463}

\newcommand{\gammawarp}{$2.00 \pm 0.01$}
\newcommand{\nhwarp}{$12.61 \pm 3.30$}
\newcommand{\scatwarp}{$2.9 \pm 1.3 \times {10}^{-3}$}
\newcommand{\inclwarp}{$51 \pm 21$}

\newcommand{\gammactorus}{$2.43 \pm 0.03$}
\newcommand{\nhctorus}{$\geq9.93$}
\newcommand{\scatctorus}{$30 \pm 6 \times {10}^{-2}$}
\newcommand{\inclctorus}{$\geq 85$}

\newcommand{\gammamytorus}{$\geq2.6$}
\newcommand{\nhmytorus}{$6.4 \pm 0.4$}
\newcommand{\scatmytorus}{$\leq5 \times {10}^{-3}$}
\newcommand{\inclmytorus}{$81.8 \pm 0.5$}

\newcommand{\gammawadaring}{$1.97 \pm 0.007$}
\newcommand{\nhwadaring}{$16 \pm 1$}
\newcommand{\scatwadaring}{$\leq 3 \times {10}^{-4}$}
\newcommand{\inclwadaring}{$82 \pm 8$}

\newcommand{\changed}\textbf

\makeatother

\usepackage{listings}

\begin{document}
\title{Physically motivated X-ray obscurer models}
\abstract{The nuclear obscurer of Active Galactic Nuclei (AGN) is poorly understood
in terms of its origin, geometry and dynamics. }{We investigate whether
physically motivated geometries emerging from hydro-radiative simulations
can be differentiated with X-ray reflection spectroscopy.}{ For two
new geometries, the radiative fountain model of Wada (2012) and a
warped disk, we release spectral models produced with the ray tracing
code XARS. We contrast these models with spectra of three nearby AGN
taken by \emph{NuSTAR} and \emph{Swift}/BAT.}{Along heavily obscured
sight-lines, the models present different 4--20~keV continuum spectra.
These can be differentiated by current observations. Spectral fits
of the Circinus Galaxy favor the warped disk model over the radiative
fountain, and clumpy or smooth torus models.}{The necessary reflector
($\NH\geq10^{25}\mathrm{cm}^{2}$) suggests a hidden population of
heavily Compton-thick AGN amongst local galaxies. X-ray reflection
spectroscopy is a promising pathway to understand the nuclear obscurer
in AGN.}
\author{Johannes Buchner\inst{1,2,3,4}\thanks{\protect\href{mailto:mailto:johannes.buchner.acad@gmx.com}{johannes.buchner.acad@gmx.com}},
Murray Brightman\inst{5}, Mislav Balokovi\'{c}\inst{6,7}, Keiichi
Wada\inst{8}, Franz E. Bauer\inst{1,3,9}, Kirpal Nandra\inst{4}}
\institute{Pontificia Universidad Católica de Chile, Instituto de Astrofísica,
Casilla 306, Santiago 22, Chile\and
Excellence Cluster Universe, Boltzmannstr. 2, D-85748, Garching, Germany\and
Millenium Institute of Astrophysics, Vicuña MacKenna 4860, 7820436
Macul, Santiago, Chile\and
Max Planck Institute for Extraterrestrial Physics, Giessenbachstrasse,
85741 Garching, Germany\and
Cahill Center for Astrophysics, California Institute of Technology,
1216 East California Boulevard, Pasadena, CA 91125, USA\and
Yale Center for Astronomy \& Astrophysics, 52 Hillhouse Avenue, New
Haven, CT 06511, USA\and
Department of Physics, Yale University, P.O. Box 2018120, New Haven,
CT 06520, USA\and
Kagoshima University, Kagoshima 890-0065, Japan\and
Space Science Institute, 4750 Walnut Street, Suite 205, Boulder, Colorado
80301}
\titlerunning{New constraints on the obscurer geometry in AGN}
\authorrunning{Buchner et al.}
\maketitle

\section{Introduction}

Most Active Galactic Nuclei (AGN) are obscured by dense gas and dust
with column densities of $\NH=10^{22\mathrange25}\mathrm{cm}^{-2}$
\citep[e.g.,][]{Ueda2003,Buchner2015}. Of this, galaxy-scale gas
provides at most a minority of this obscuration at moderate column
densities $\NH=10^{22\mathrange23.5}\mathrm{cm}^{-2}$ \citep[e.g.,][]{Buchner2017a}.
The predominant obscurer is thought to be a parsec-scale circum-nuclear
structure \citep[see e.g.][for reviews]{Antonucci1993,Netzer2015,RamosAlmeida2017}.
However, the geometry, sub-structure and physical origin of this component
is still unclear and heavily debated.

Already early modelling efforts \citep[see, e.g.,][and references therein]{Matt2000a}
revealed that the unresolved X-ray continuum of AGN is composed of
a power law source (with a high-energy cut-off seen in some sources)
that is photo-electrically absorbed. The absorbing screen primarily
suppresses soft photons, bending the power law downwards towards soft
energies. This suppression reveals additional, secondary components
interpreted as ionised and cold gas reprocessing and reflecting the
primary power law emission. The ionised reprocessing, also called
the `warm mirror', can be interpreted as Thomson scattering off
ionised, stratified volume-filling gas, potentially in the narrow-line-region
\citep[e.g.,][and references therein]{Turner1997a,Bianchi2006}. This
spectral feature is to first order a copy of the intrinsic power law,
but with the normalisation approximately $0.1\mathrange10\%$ of the
intrinsic emission, and is commonly seen in obscured AGN \citep[e.g.,][]{Rivers2013,Buchner2014,Brightman2014,Ricci2017a}.

The interaction in cold reprocessing is more complex. About a third
of all AGN have line-of-sight (LOS) column densities $N_{\mathrm{H}}$
in excess of the Compton-thick limit ($\NH=\sigma_{T}^{-1}\approx1.5\times10^{24}\NHunit$;
\citealp[e.g.,][]{Buchner2015,Ricci2016}). This implies that Compton
scattering is an important process. X-rays from the corona recoil
off distant, cold, dense gas, lose energy and change direction. The
Compton scattered spectrum depends on the geometry and density of
the reflecting cold matter. The reflection was first modelled as simple,
semi-infinite slabs \citep[e.g.][]{Magdziarz1995,Matt1999a}. Slabs
are designed to mimic the accretion disk and produces a very hard
spectrum (negative effective photon index), decorated with fluorescent
lines such as Fe~K$\alpha$ and absorption edges. An important feature
is excess emission around $10\mathrange50\mathrm{keV}$, the \emph{Compton
hump}. As knowledge and capabilities in spectral and spatial resolution
progressed, more realistic modelling was needed, allowing refined
parameter constraints and interpretations \citep[e.g., models by][]{MurphyYaqoobMyTorus2009,Brightman2011a,Ikeda2009}.
Today, the obscurer geometry is still uncertain. This is partly because
the observed X-ray emission can be contaminated by nearby X-ray binaries,
supernova remnants and ultra-luminous X-ray sources \citep[e.g.,][]{Arevalo2014},
and because extended emission can contribute significantly to the
Fe~K$\alpha$ flux \citep[e.g.,][]{Bauer2014}. Even with these taken
into account, detailed study of the nearby Compton-thick AGN \object{NGC~1068}
reveals multiple cold reflectors of different column densities \citep{Bauer2014}.
Also, there appears to be significant diversity between sources \citep[e.g.,][]{Balokovic2018}.
Other studies find that the reflection in heavily obscured sources
is best-fitted by a face-on orientation of a torus-shaped geometry
\citep[e.g.][]{Balokovic2014,Gandhi2014}. Finally, observations of
variability in measured column density \citep[e.g.,][]{Risaliti2002,Markowitz2014}
indicates that the obscurer is clumpy and subsequently, the first
clumpy models have been developed \citep[e.g.,][]{Nenkova2008a,Furui2016,Liu2014,Buchner2019a,Tanimoto2019}.
The applied X-ray obscurer models are perhaps the simplest geometries
for a unified obscurer. However, it remains unclear which specific
physical mechanisms maintain the high covering factors of these geometries
and where the obscuring gas originates.

The above considerations indicate that our understanding of the obscurer
geometry is still incomplete, in part due to a dearth of self-consistent
absorber/reflector models with physical motivations. In the last decade,
several theoretical models based on radiative hydro-dynamic simulations
have been developed, which attempt to explain the origin and stability
of the obscurer. The high covering is achieved either by bringing
in host galaxy scale gas from random orientations \citep[e.g.,][]{Hopkins2011,Gaspari2015},
by black hole and star formation feedback processes in the inner hundred
parsec \citep[e.g.,][]{Wada2012,Schartmann2018}, or by the black
hole accretion system in isolation \citep[e.g.,][]{Chan2016,Dorodnitsyn2017,Williamson2018,Moscibrodzka2013}.
However, such models have not yet been tested against X-ray reflection
spectroscopy. We investigate the spectral signatures of different
physically motivated geometries, and whether they can be differentiated
by current X-ray instruments. In section~\ref{sec:methodology},
we develop an open-source Monte Carlo code that can be used to simulate
the irradiation of arbitrary grid geometries and produce X-ray spectra.
We investigate two new X-ray spectral predictions based on physical
models, presented in section~\ref{sec:Models}. The geometries were
dynamically created in the environment of super-massive black holes
by effects like gravity and radiation pressure. Section~\ref{sec:Results}
presents the emerging X-ray spectra, and section~\ref{sec:Comparison-to-observations}
validates them against X-ray observations. Section~\ref{sec:Discussion}
discusses the physical implications of the results and suggests future
observational tests.

\section{Methodology}

\subsection{Simulations}

\label{sec:methodology}

Monte Carlo simulations are used to compute the emerging X-ray spectrum
for several geometries of interest. For this, we developed a modular,
Python-based simulation code \XARS{} (X-ray Absorption Re-emission
Scattering, \citealp{Buchner2019a}), which is publicly available
at \url{https://github.com/JohannesBuchner/xars}. The simulation
method is described in detail in \citet{Brightman2011a}. Briefly,
a point-source emits photons isotropically from the centre of the
obscurer geometry. Photo-electric absorption, Compton scattering and
line fluorescence are simulated self-consistently as photons pass
through matter, altering the direction and energy of the photons or
absorbing them. For simplicity and to compare with other works, solar
abundances from \citet{Anders1989} are assumed with cross-sections
from \citet{Verner1996}. \XARS{} computes the Green's functions
at each defined energy grid point and viewing angle, onto which an
input photon spectrum, e.g. a powerlaw, can be applied, and transformed
into an \xspec{} grid model.

Users of\texttt{ }\XARS{} specify the geometry of the obscurer by
implementing how far, starting from a given location and direction,
photons can propagate through the medium. For this work we additionally
implemented photon propagation through 3-dimensional density grids,
such as those produced from hydrodynamic simulations. Our parallelised
and optimised C implementation is based on the 3D Digital Differential
Analyser \citep{4056861} and publicly available within the \texttt{LightRayRider}
library \citep{Buchner2017} at \url{https://github.com/JohannesBuchner/LightRayRider}
and can easily be used with \XARS{} (examples are included in the
\XARS{} documentation). Digital Differential Analysers are a simple
but efficient technique to traverse grids, and compute the length
before a new grid line is encountered. Assuming constant-density grid
cells, this allows us to integrate the column density along the photon
path efficiently. The exact technique is described in Appendix~\ref{sec:3DDDA}.
This enables the computation of X-ray spectra emerging from arbitrary
density distributions. 

\subsection{Observations}

\label{subsec:Observations}

After presenting the new spectral models in the following sections,
preliminary comparisons to observations are made. A rigorous comparison
of careful spectral fits to an unbiased sample is left for future
work. Here, we test the plausibility of the models on three sources:
Circinus Galaxy, NGC~424 and ESO 103-G03. These are bright and well-studied
examples of nearby AGN that demonstrate a variety of spectral features
of interest in our study. The \emph{NuSTAR} spectra of the three sources
was previously analysed in \citet{Buchner2019a}. All three show prominent
reflection on top of a obscured continuum, but their accretion rate
and line-of-sight obscuration is diverse. They were targeted by \emph{NuSTAR}
in single-epoch observations (ObsIds: 60061007002, 60002039002, 60061288002)
for $10\mathrange60\,\mathrm{ks}$. Spectra were extracted using the
NuSTAR data processing tools in the same manner as in \citet{Brightman2015}.
The \emph{NuSTAR} spectra are binned to 20 counts per bin and $\chi^{2}$
statistics adopted, which is justified in these high count spectra.

Of our three sources, the closest and best understood is the Circinus
Galaxy. One of the model geometries we consider was specifically developed
for this galaxy. For this reason, we include a detailed spectral analysis
of this source. Within the point spread function of current hard X-ray
instruments, extended emission is present \citep[e.g.,][]{Bauer2014,Fabbiano2018},
and below $10\,\mathrm{keV}$ components from star formation and X-ray
binaries are important \citep{Marinucci2013,Arevalo2014}. For this
reason, we include these components in a joint fit. To focus on the
high-energy shape, we consider the \emph{NuSTAR} spectrum from $8\mathrange75\mathrm{keV}$
and add \emph{Swift}/BAT spectra from the 105 month data release \citep{Oh2018}.
For the other two sources, we only make visual comparisons with some
models above 8 keV as a first-order validation, while more detailed
analyses are left for a future publication.

\section{Geometries}

\label{sec:Models}

Realistic obscurer models need to explain the high obscured fraction
in the AGN population. This requires sustaining large covering factor
around the X-ray source \citep[see, e.g.,][]{Lawrence2010,Wada2015,Schartmann2018}.
Below we present two physically motivated X-ray obscurer models: A
warped disk and the radiative fountain model. In the following section,
we discuss the resulting spectra.

\subsection{Wada's radiative fountain model\label{subsec:Wada's-hydro-radiative-fountain}}

\begin{figure}
\begin{centering}
\includegraphics[viewport=0bp 0bp 279bp 279bp,width=0.7\columnwidth]{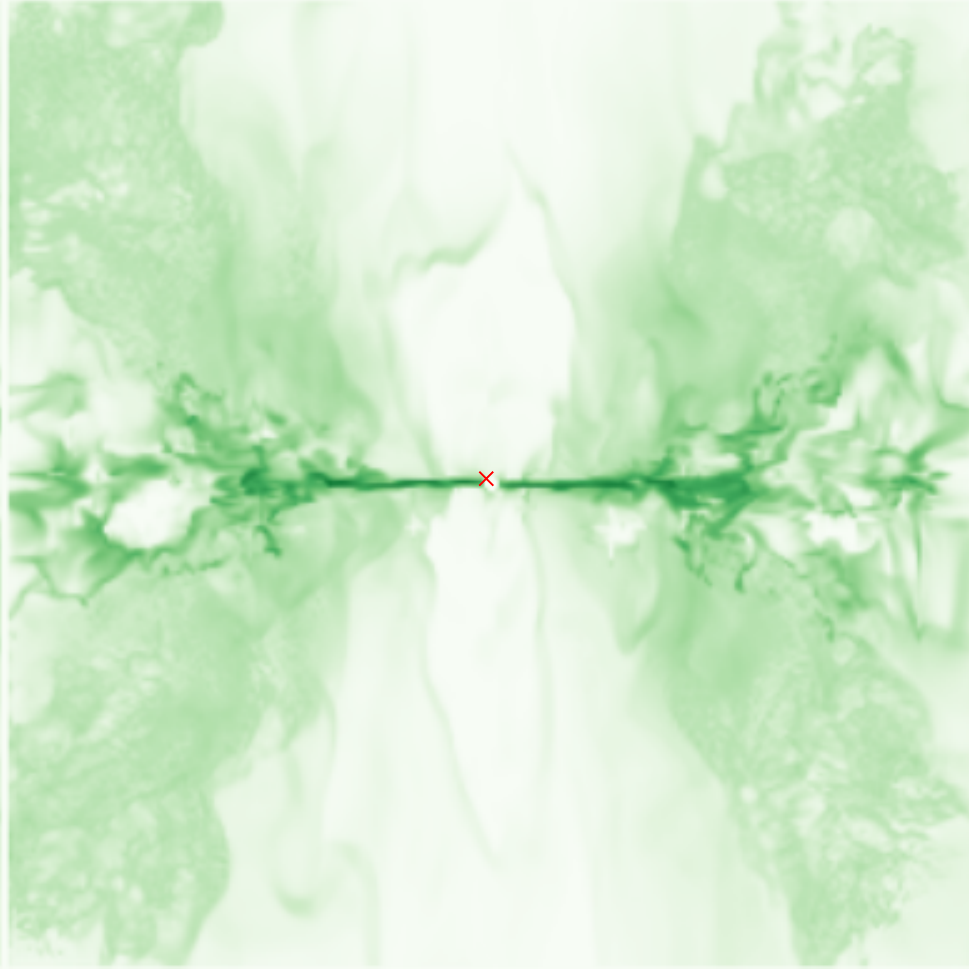}
\par\end{centering}
\caption{\label{fig:WadaGeometry-1}Geometry of the radiation-driven fountain
model of \citet{Wada2016}. A cross-section of the density grid is
shown with the X-ray emitting corona indicated by the red cross. The
side length of the simulation region is $32\mathrm{pc}$.}
\end{figure}

\citet{Wada2012} developed a model in which the obscurer is a radiation
pressure-driven polar outflow from a black hole accreting within a
thick, parsec-scale disk. The outflow falls back akin to a fountain
(see Figure~\ref{fig:WadaGeometry-1}) and forms a torus-like shape
producing suitably high covering factors \citep{Wada2015}. In \citet{Wada2016},
supernova feedback is also included. The evolution of this system
was computed with three-dimensional radiation hydrodynamic simulations
on a $256^{3}$ cell grid spanning a domain of $(32\mathrm{pc})^{3}$.
As a representation of this model, we use the \citet{Wada2016} geometry
from their final simulation time step. Figure~\ref{fig:WadaGeometry-1}
shows the geometry for a black hole mass of $M_{{\rm BH}}=2\times10^{6}M_{\odot}$
and 20\% Eddington rate, suitable for comparison to Circinus \citep[see][for comparison with infrared SEDs, atomic/molecular gas, and ionized gas observations, respectively]{Wada2016,Wada2018,Izumi2018,Wada2018a}.
The model has a Compton-thick LOS column density only under edge-on
viewing angles \citep[consistent with the infrared analysis of Circinus by][]{Wada2016}.

The X-ray spectrum under this geometry is computed by placing the
X-ray corona at the centre of the simulations density grid (see Figure~\ref{fig:WadaGeometry-1}).
We use \XARS{} to irradiate the geometry with $5\times10^{9}$ photons,
and capture the energy and direction of X-ray photons escaping to
infinity. The direction is used to divide photons by viewing angle
in two steps. First, the LOS column density ($\NH$) from the corona
to infinity in that direction is computed, which defines the first
binning axis. Secondly, we further sub-divide by azimuthal angle (face-on
to edge-on). This scheme \citep[also in][]{Buchner2019a} has the
benefit of separating out the most important observable (LOS $\NH$),
and allowing varying the LOS $\NH$ and viewing angle independently.

\begin{figure}
\begin{centering}
\includegraphics[width=0.7\columnwidth]{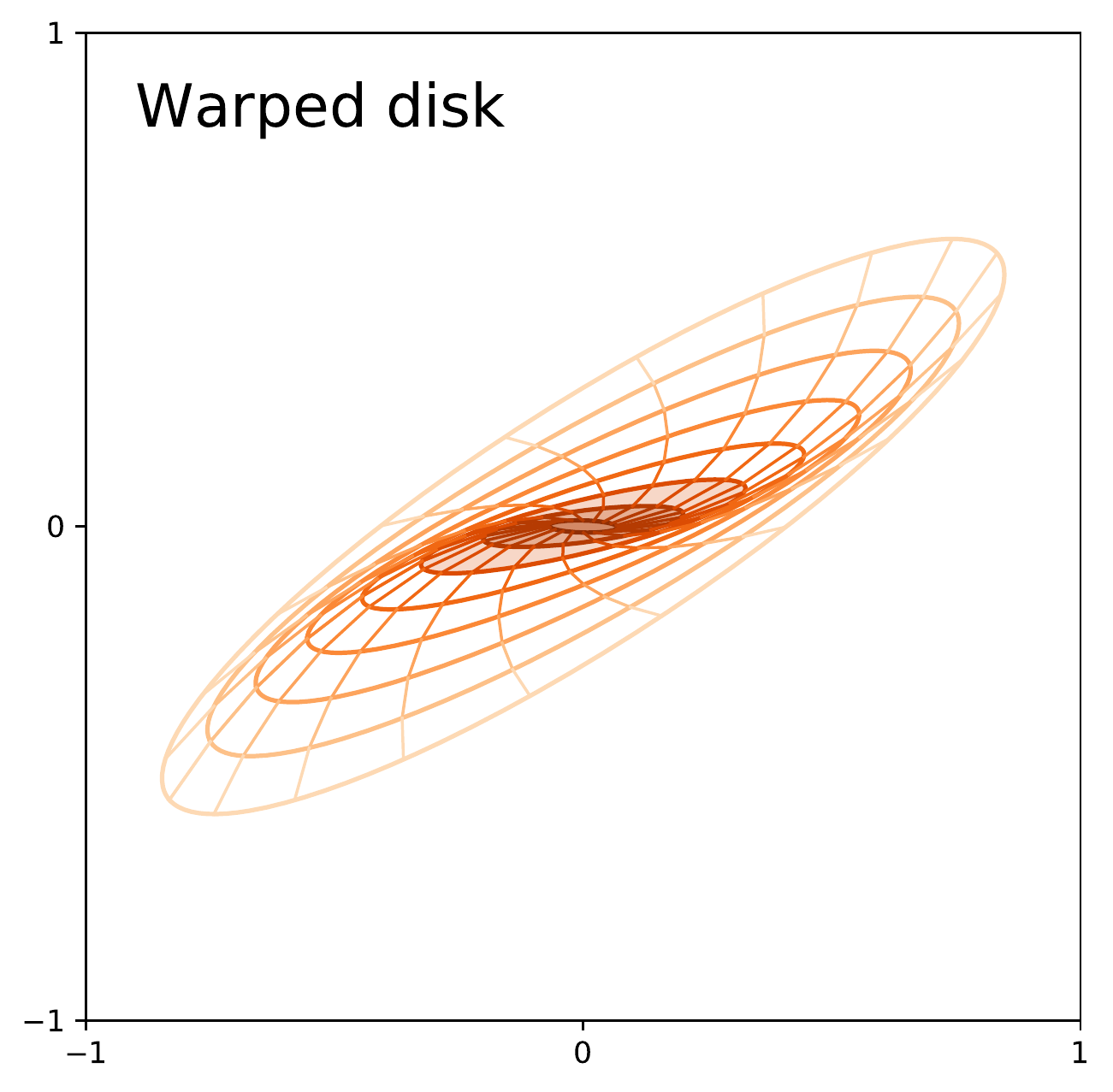}
\par\end{centering}
\caption{\label{fig:warpeddisk-geometry-1}Warped disk geometry, viewed edge-on.
Near the centre the disk is only mildly warped (shaded inner-most
two annuli, $r_{\text{warp}}\sim0.25$), and viewing angles are defined
relative to there. The outer disk annuli (up to radius $r_{\text{warp}}=1$)
are warped more strongly and form a wall, suitable for Compton reflection.
For our simulation, we set up a grid with $256^{3}$ cells and assign
cells intersecting with this shape a Compton-thick column density.}
\end{figure}

\subsection{Warped disk\label{subsec:Warped-disk}}

Warped, tilted and twisted disks have been proposed as the source
of (Compton-thick) obscuration \citep[e.g.,][]{Lawrence2010}. Non-planar
accretion can cause precessing concentric rings of accretion disks
\citep{Petterson1977}. Additionally, the disk emits normal to the
local surface, but the received radiation pressure from the centre
is misaligned and causes torque, and thus twisted disks \citep{Petterson1977a}.
The shape distortion is limited by viscosity, as gas moves radially
inward.

Water maser disks are one type of such warped disks known to exist
in several AGN. Strongly emitting disk clumps trace out edge-on warped
disk structures, e.g. in Circinus \citep{Greenhill2003}. Recently,
\citet{Jud2017} developed an infrared model for the warped disk.
They demonstrate good fits to infrared photometry and can naturally
explain the offset between point source and disk component seen in
the highest resolution infrared VLTI observations \citep{Tristram2014}.
Water maser disk systems are always found to be heavily obscured \citep{Greenhill2008,Masini2016},
indicating that the Compton-thick obscurer either is the warped maser
disk itself or at least shares the same plane of orientation. 

We now simulate such a warped disk obscurer. \citet{Pringle1996}
describes the geometry of a warped disk as 
\begin{eqnarray*}
x & = & r\cdot(\sin\phi\cos\gamma\cos\beta+\cos\phi\sin\gamma)\\
y & = & r\cdot(\sin\phi\sin\gamma\cos\beta-\cos\phi\cos\gamma)\\
z & = & r\cdot(-\sin\phi\sin\beta)
\end{eqnarray*}
with $\gamma(r)=\sqrt{r}$ and $\beta(r)=\sin(\gamma(r))/\gamma(r)$,
following \citet{Maloney1996}. The parametrisation essentially describes
a disk ring at radius $r$ across angles $\phi\in[0,2\pi]$. We set
up a computation grid of $256^{3}$ cells and assign a column density
of $\NH=10^{25}\mathrm{cm}^{-2}$ to all grid cells intersecting the
warped disk, as well as the cells just below to produce a thickness
of at least two grid cells everywhere. 

We set up various warp strengths. We let the radii $r$ range from
zero (the centre) up to $r_{\text{warp}}=1$, $1/2$, $1/4$ or $1/16$.
The case of a strong warp ($r_{\text{warp}}=1$) is illustrated in
Figure~\ref{fig:warpeddisk-geometry-1} and features a prominent
wall. For $1/4$ and $1/16$, only the flat thin (unwarped) inner
disk remains, illustrated as the inner annuli of Figure~\ref{fig:warpeddisk-geometry-1}. 

\begin{figure*}
\begin{centering}
\includegraphics[width=0.5\textwidth]{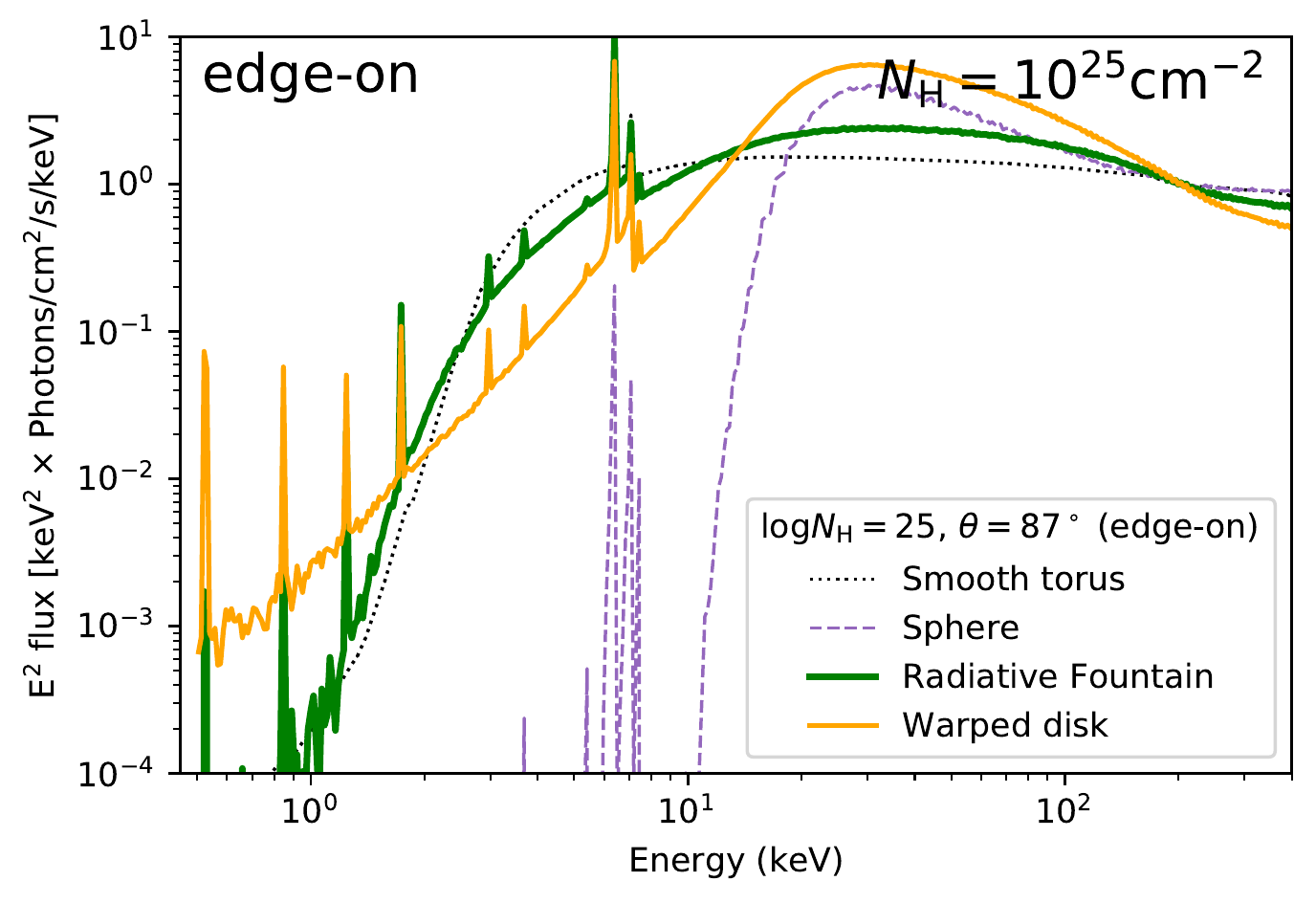}\includegraphics[width=0.5\textwidth]{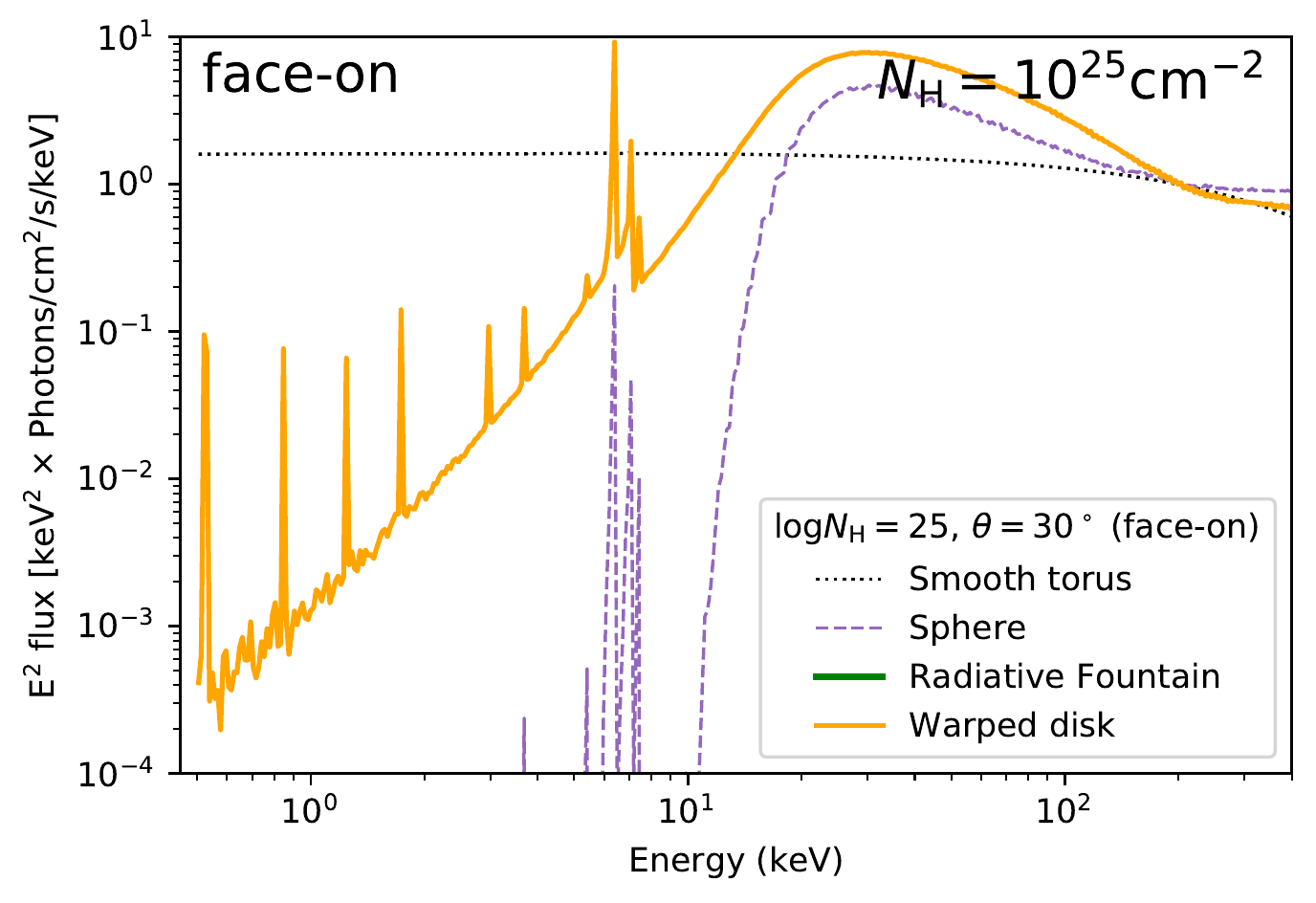}
\par\end{centering}
\begin{centering}
\includegraphics[width=0.5\textwidth]{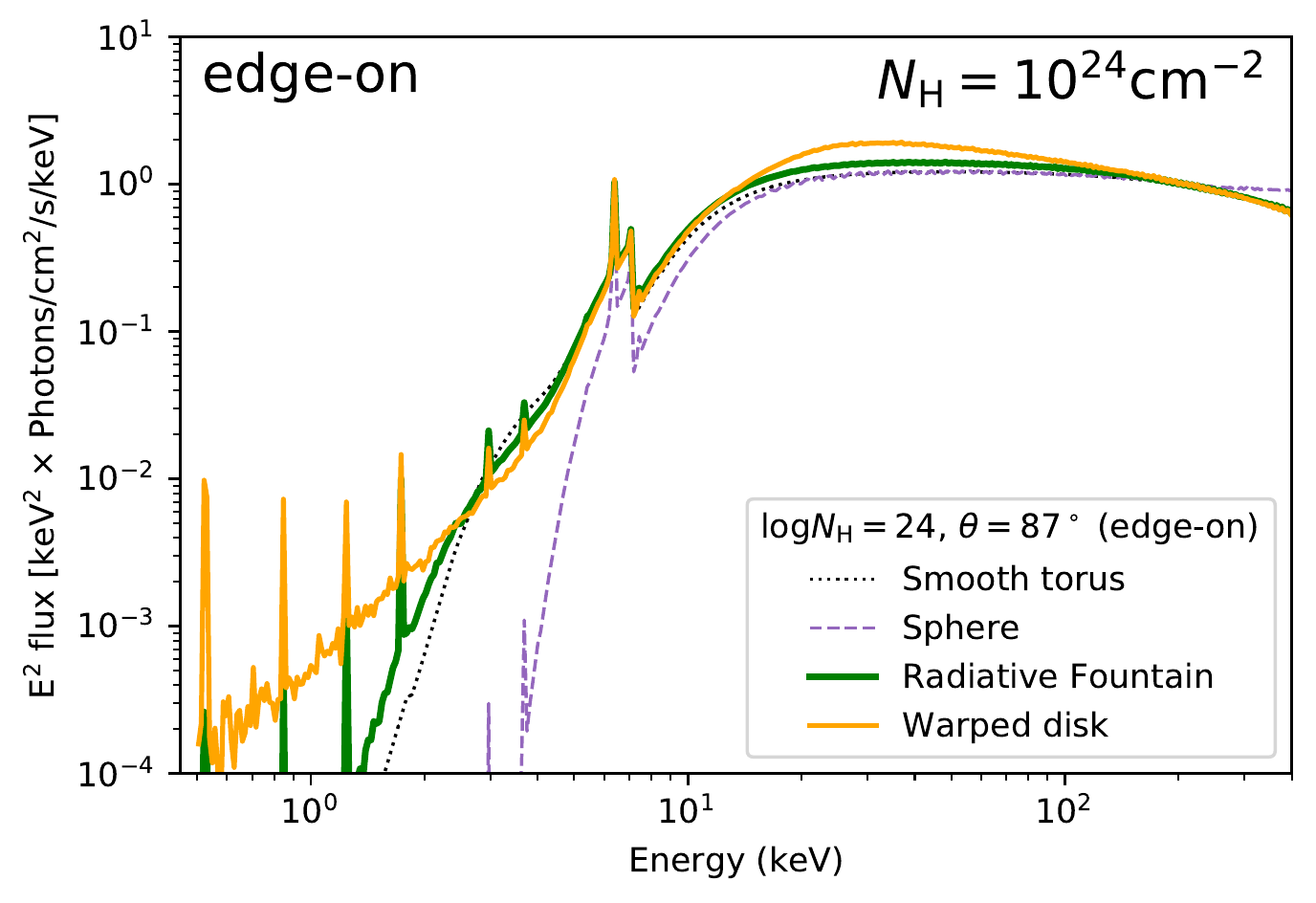}\includegraphics[width=0.5\textwidth]{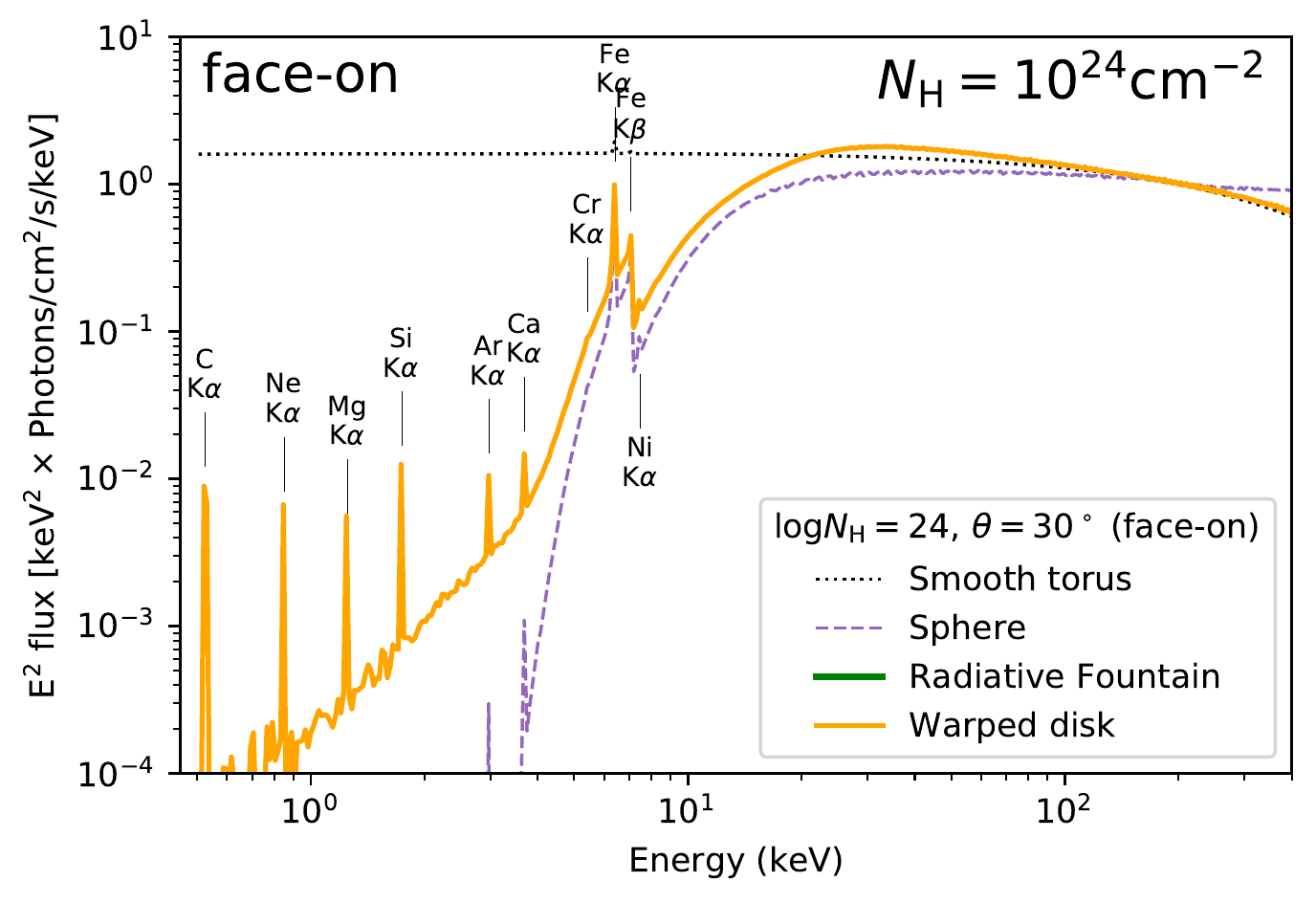}
\par\end{centering}
\begin{centering}
\includegraphics[width=0.5\textwidth]{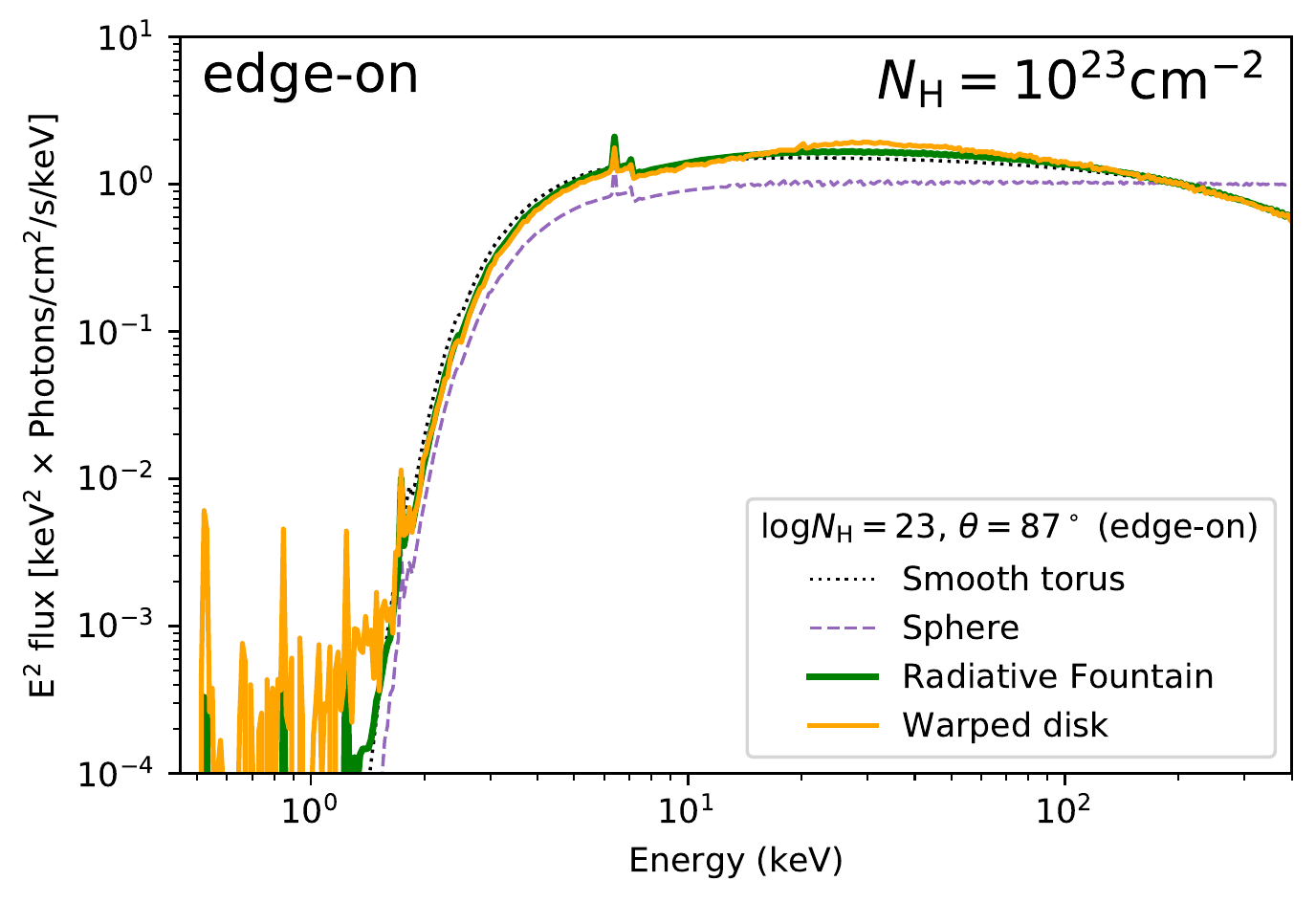}\includegraphics[width=0.5\textwidth]{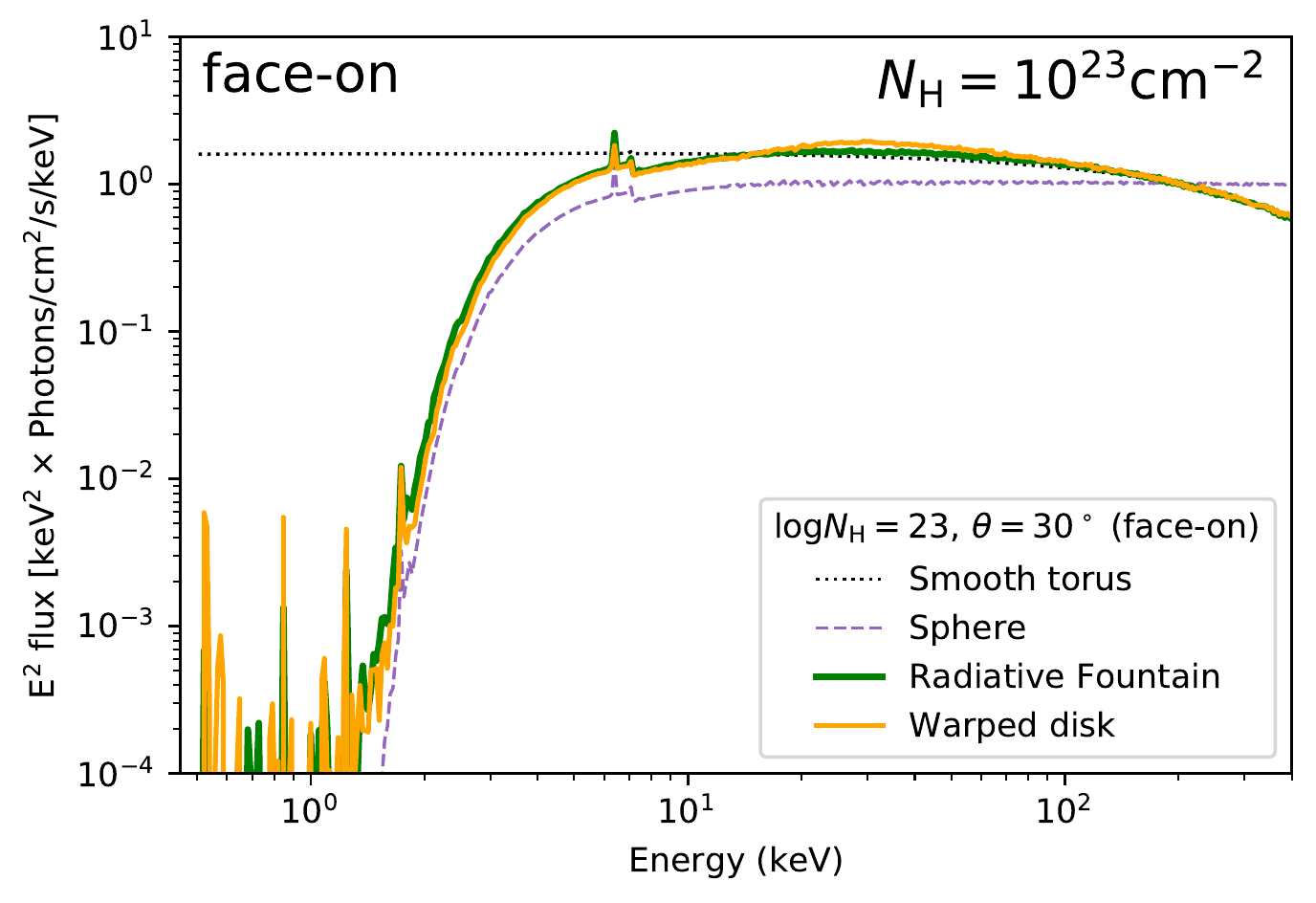}
\par\end{centering}
\caption{\label{fig:modelspectra}Model spectra of the radiative fountain model
(green) and warped disk model (orange). The rows show different line-of-sight
column densities. The left/right column shows a nearly edge-on/face-on
view of the obscurer. For comparison, a spherical obscurer \citep{Brightman2011a}
and smooth torus \citep{MurphyYaqoobMyTorus2009} are included as
purple dashed and black dotted curves, respectively. In low column
densities (bottom row) the predictions are similar. In the right column,
the radiative fountain predictions are absent because near-Compton-thick
column densities only occur in the equatorial plane. Similarly, the
torus is unobscured from this viewing angle and shows the input powerlaw.
The models differ most in the $2\mathrange20\mathrm{keV}$ range under
high obscuring column densities (top left panel). The warped disk
produces the strongest emission near $20\mathrm{keV}$ among the models.
Models are normalised at $200\,\mathrm{keV}$. The center right panel
annotates the included fluorescent lines.}

\end{figure*}

\section{Results}

\subsection{Continuum shapes}

\label{sec:Results}

\begin{figure}[h]
\begin{centering}
\includegraphics[width=1\columnwidth]{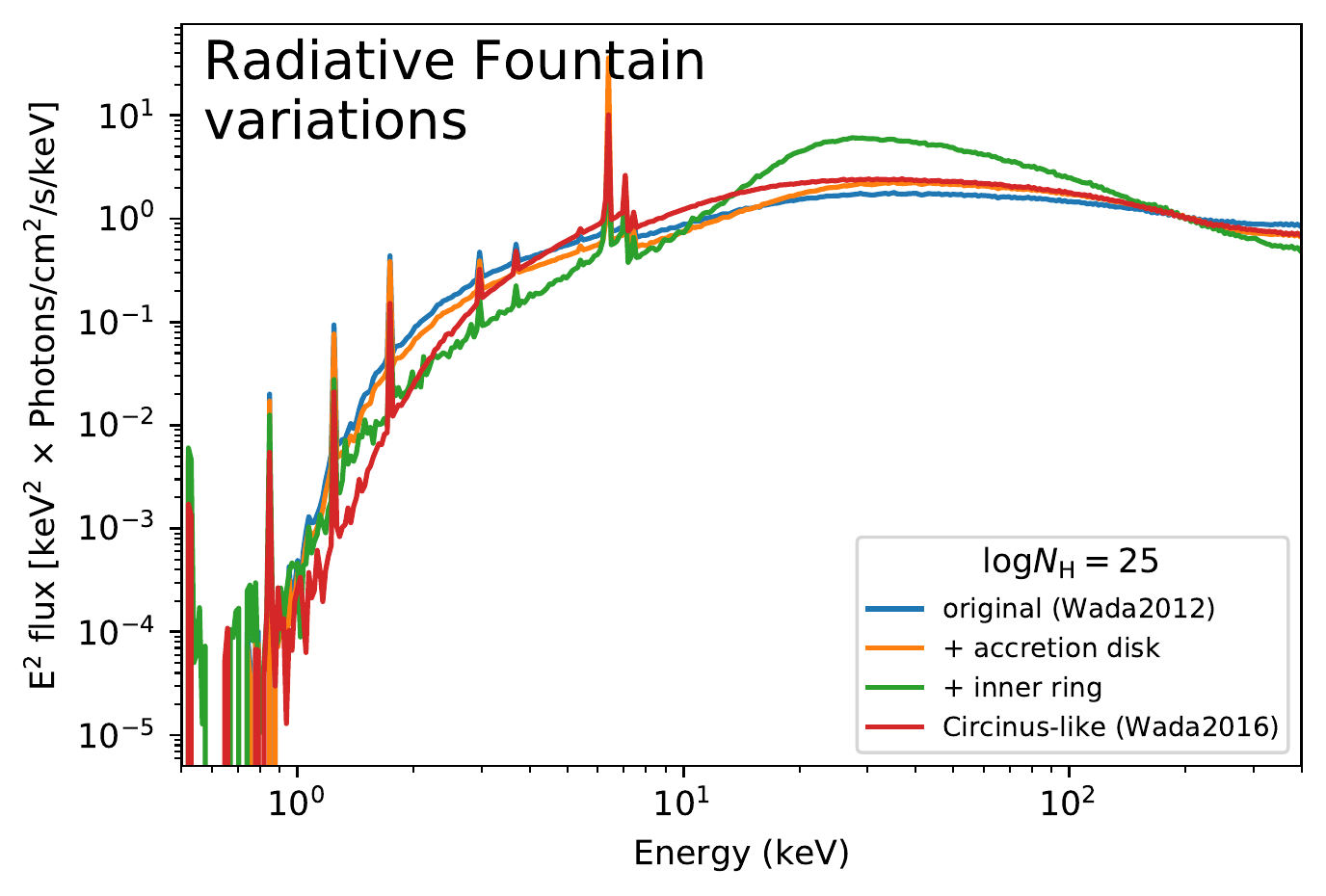}
\par\end{centering}
\caption{\label{fig:modelspectra-wada}Variations of the radiative fountain
model spectra. The \citet{Wada2012} and \citet{Wada2016} models
(blue and red, respectively) differ in black hole mass and supernova
feedback. The orange curve includes reflection from an accretion disk.
The green curve includes a Compton-thick ring around the central X-ray
source, which leads to stronger emission in the $10\mathrange50\,\mathrm{keV}$
energy range than the original model (blue curve). Models are normalised
at $200\,\mathrm{keV}$.}
\end{figure}
\begin{figure}[h]
\begin{centering}
\includegraphics[width=1\columnwidth]{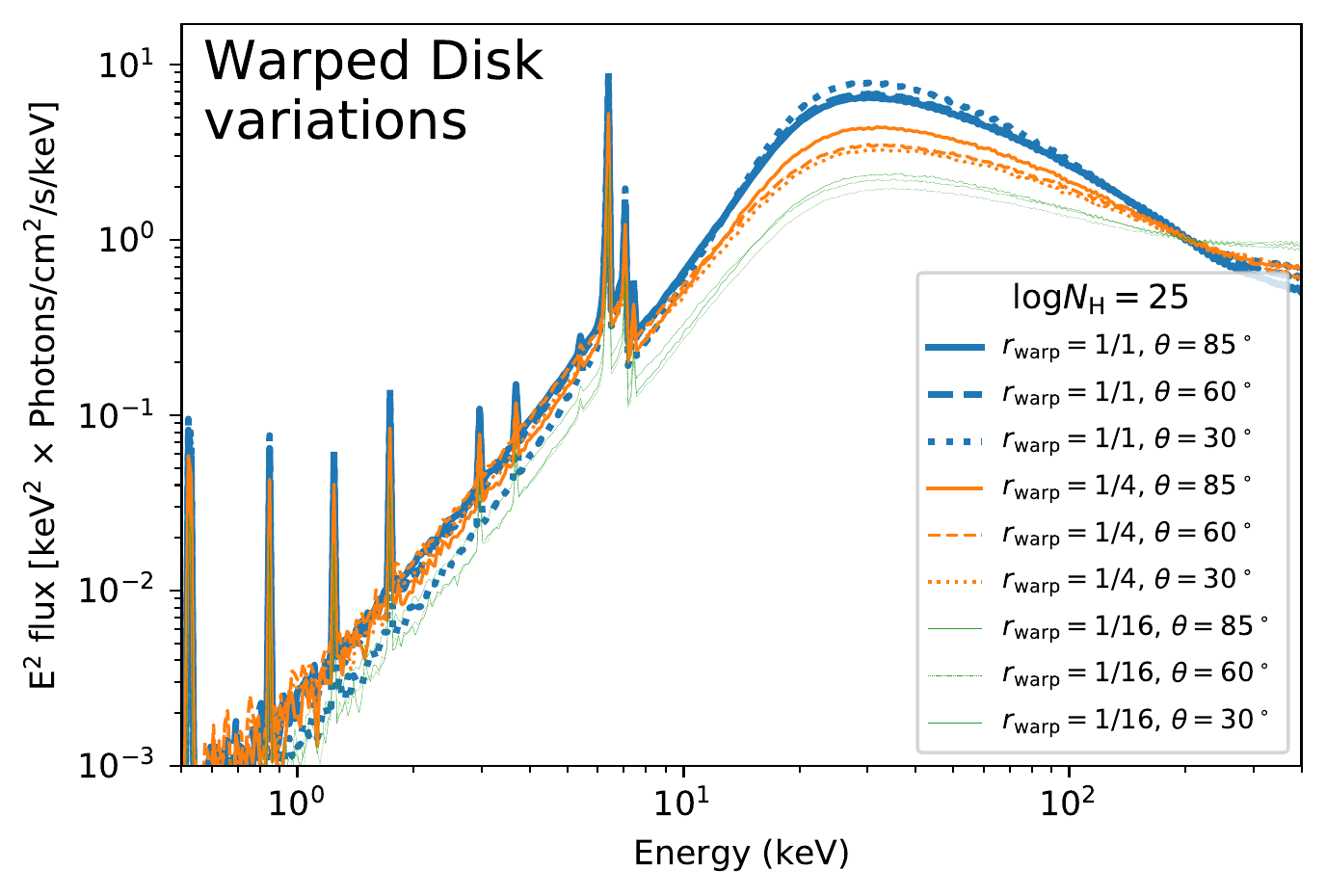}
\par\end{centering}
\caption{\label{fig:modelspectra-warp}Variations of the warped disk model.
Variations of the inclination angle relative to the innermost annuli
are represented by solid, dashed and dotted lines. Variations of the
extent of the warp (see Figure~\ref{fig:warpeddisk-geometry-1})
are represented by line strengths. The thickest lines (strong warps)
exhibit the strongest emission in the $20\mathrange30\,\mathrm{keV}$
range. Models are normalised at $200\,\mathrm{keV}$.}
\end{figure}

Examples of obscured spectra are presented in Figure~\ref{fig:modelspectra}.
The input photon spectrum is a cut-off powerlaw $E^{-\Gamma}\times\exp(-E/E_{\mathrm{cut}})$
with photon index $\Gamma=2$ and exponential cut-off $E_{\mathrm{cut}}=400\mathrm{keV}$.
The panels of Figure~\ref{fig:modelspectra} plot the emerging $E^{2}\times$photon
spectrum, split by inclination and LOS column density. For orientation,
we include two models based on simple geometries in Figure~\ref{fig:modelspectra}.
The sphere model of \citet{Brightman2011a} embeds the X-ray point
source in a constant-density sphere. This model completely suppresses
photons below $10\,\mathrm{keV}$ in Compton-thick sight-lines, except
for Fe~K fluorescence. The smooth torus model of \citet{MurphyYaqoobMyTorus2009}
surrounds the X-ray point source with a constant-density torus, with
an opening angle of 60°. Its soft emission is dominated by Compton
reflection off the torus. In moderately obscured sight-lines (bottom
panels of Figure~\ref{fig:modelspectra}), all models make similar
predictions. The sphere model differs slightly from the others, as
it does not include an exponential cut-off. 

The models differ in the strength of the Compton hump near $10\mathrange50\,\mathrm{keV}$.
The warped disk model shows the strongest emission, while the radiative
fountain model has a weak Compton hump. This is likely related to
the high covering factor of high-density material in the warped disk
that is seen without further obscuration by the observer. The two
models bracket the sphere model in terms of the Compton hump strength.
Additionally, the hump shapes are different in their turn-over.

The models differ most strongly in the $2\mathrange20\,\mathrm{keV}$
energy range under Compton-thick LOS (e.g., top left panel). Here
the transmitted powerlaw continuum is suppressed, revealing the Compton
reflected components and fluorescent lines. In both of our two models,
soft photons escape under all viewing angles through Compton scattering.

The spectral slopes in the $2\mathrange20\,\mathrm{keV}$ energy range
are diverse. The spherical obscurer shows the steepest slope among
the models. The smooth torus and radiative fountain models show a
smooth exponential turn-over, at lower energies than the sphere geometry.
In contrast, the warped disk model presents a powerlaw-like continuum
in the $2\mathrange20\,\mathrm{keV}$ energy range with photon index
$\Gamma\approx0.5$. The warped disk produces the strongest emission
near $20\,\mathrm{keV}$ among the models. Its Compton hump is widest
in the radiative fountain model. 

We explore these differences further in Figure~\ref{fig:modelspectra-wada}
and \ref{fig:modelspectra-warp}. Figure~\ref{fig:modelspectra-wada}
shows the spectra from the models of \citet{Wada2012} and \citet{Wada2016},
which differ primarily in that the latter uses a ten times smaller
black hole mass and the presence of supernova feedback, both chosen
to match the Circinus Galaxy. The X-ray spectral predictions are however
of similar shape, characterised by a smoothly bending reflection spectrum
in the $2\mathrange20\,\mathrm{keV}$ energy range. The resulting
spectra are very similar across viewing angles, if the LOS column
density are kept constant. However, Compton-thick sight-lines, as
shown in Figure~\ref{fig:modelspectra-wada}, only exist in equatorial
views.

\subsection{Model variations}

Reflection close to the X-ray corona can alter the spectral shape.
The accretion disk is a known reflector, and can be modelled by a
semi-infinite slab. We replace the input powerlaw source with a powerlaw
source and its disk-reflected, angle-averaged spectrum. The disk model
spectrum was computed with \XARS{} and presented in the appendix
of \citet{Buchner2019a}, and is consistent with the pexmon disk model
\citep{pexmon}. The effect of the accretion disk model can be seen
by comparing the orange curve in Figure~\ref{fig:modelspectra-wada}
with the blue curve. The model with accretion disk (orange) shows
a higher Compton hump than the model without (blue). However, the
difference is less than a factor of 2. This may be because the disk
adds reflection, but does not hinder the powerlaw emission from the
X-ray source over half of the sky.

Higher covering factor obscuring structure at small scales have more
drastic changes. We experiment how the fountain model would need to
be minimally changed to behave like the warped disk model. At the
same time, the ultra-violet radiation pressure of the model should
remain unaffected. A completely ionised structure between accretion
disk and $<0.1\mathrm{pc}$ scales may be a viable solution. This
could be similar to the dense, inner shielding gas of wind simulations
by \citet{Gallagher2015}, and could also be interpreted as a sub-parsec
scale warped disk, or the broad line region. For simplicity, we achieve
a toroid with a opening angle of approximately 60°, by setting the
eight spaxels closest to the X-ray emitter in the equatorial plane
to Compton-thick column densities. The emerging spectrum is shown
in green in Figure~\ref{fig:modelspectra-wada}. The spectral shape
is dramatically different, showing a much stronger Compton hump. It
is very similar to the warped disk model (see top left panel of Figure~\ref{fig:modelspectra}).

Variations of the warped disk models are shown in Figure~\ref{fig:modelspectra-warp}.
The most important parameter is the extent of the warp (increasing
with line thickness). For stronger warps (see visualisation in Figure~\ref{fig:warpeddisk-geometry-1}),
the increased area available for Compton reflection strengthens the
Compton hump emission near $20\,\mathrm{keV}$. The viewing angle
(defined relative to the innermost annuli) has a minor effect, because
of the twisting shape of this geometry.

We publish the spectra as \xspec{} table models at \url{https://github.com/JohannesBuchner/xars}.

\subsection{Plausability of models}

\label{sec:Comparison-to-observations}

The model diversity shown in the previous section indicates that in
the Compton-thick regime, hard X-ray spectra can differentiate model
geometries. As outlined in §\ref{subsec:Observations}, we explore
the plausibility of our models by projecting them through the \emph{NuSTAR}
response and comparing against a few, high-quality observations. Typical
AGN source parameters (a powerlaw continuum with photon index $\Gamma=2$)
are assumed, and the line-of-sight obscuration is varied. As contaminating
components all decline steeply with energy, therefore we normalise
each model with the data at the high energy end and accept models
that do not overpredict the data.

\begin{figure}
\begin{centering}
\includegraphics[width=1\columnwidth]{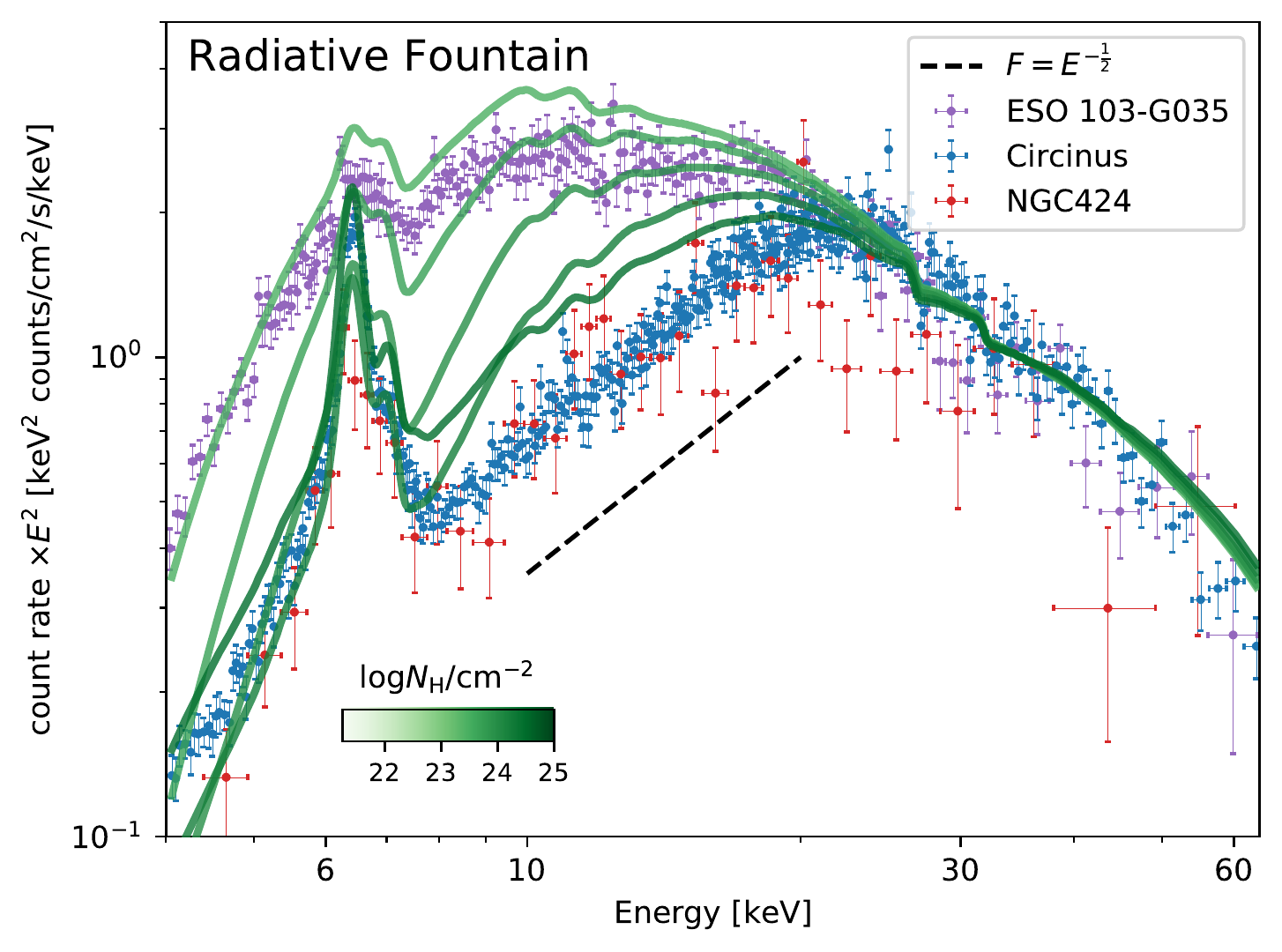}
\par\end{centering}
\begin{centering}
\includegraphics[width=1\columnwidth]{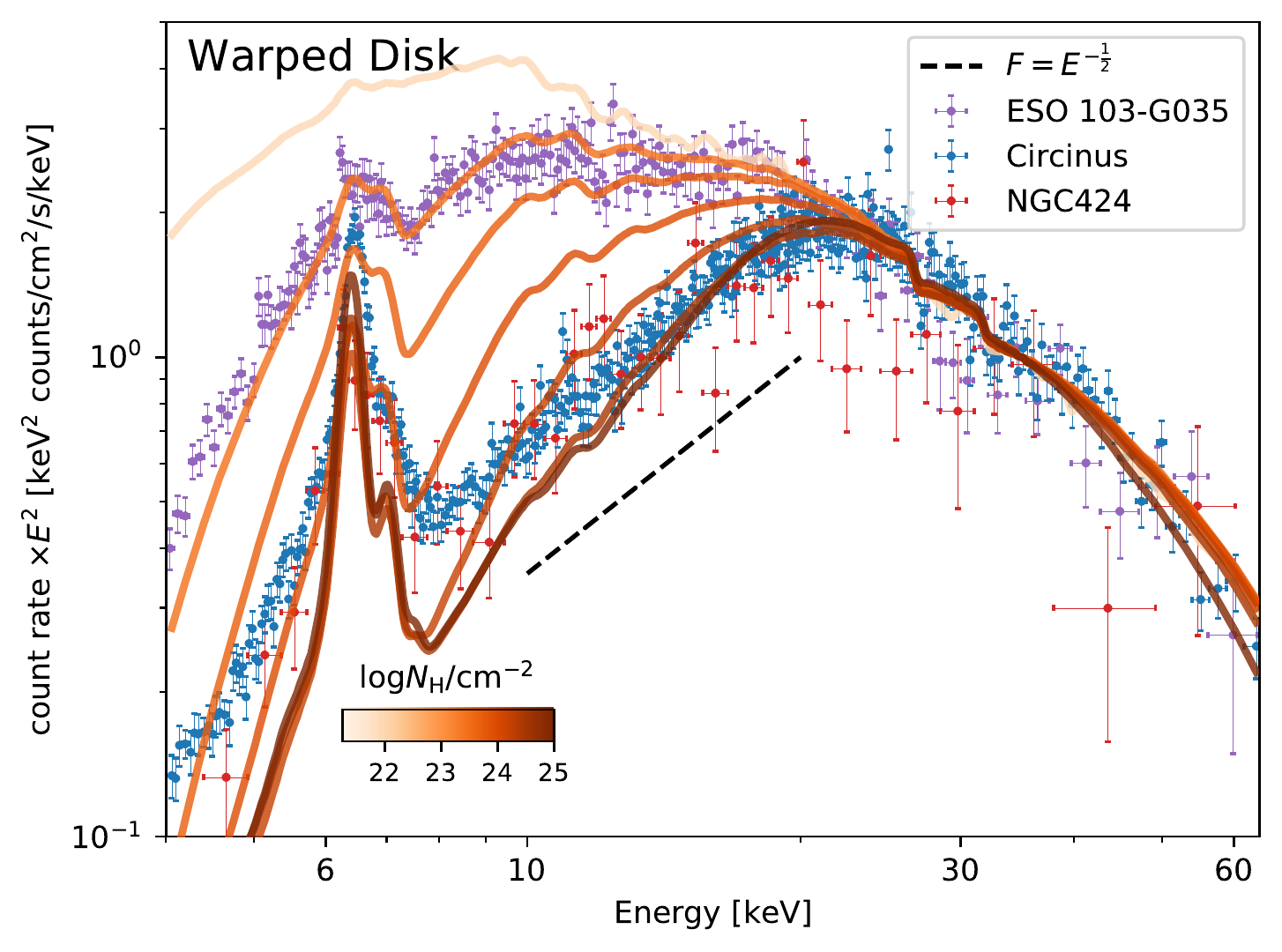}
\par\end{centering}
\caption{\label{fig:data}Comparison of radiative fountain model (\emph{top
panel}) and warped disk model (\emph{bottom}) to three observed spectra
of AGN (error bars). Each panel shows one fixed model geometry. Model
instances with photon index $\Gamma=2$ and LOS column densities of
$\log\NH=20,23.5,23.75,24,24.25,24.5,26$ (top to bottom curves) are
folded through the three instrument responses (near-overlapping groups
of curves). Data and models are normalised at $35\,\mathrm{keV}$.
\emph{Top panel}: The radiative fountain models show an excess near
$10\mathrm{\,keV}$ compared to the data. \emph{Bottom panel}: The
warped disk model follows the ESO~103-G035 data closely, and also
reproduces the Circinus data. The models underproduce the data at
$4\mathrange8\,\mathrm{keV}$.}
\end{figure}

Figure~\ref{fig:data} shows three \emph{NuSTAR} spectra of nearby
AGN. The high-quality data reveals the reflector in great detail.
The top panel of Figure~\ref{fig:data} compares spectra from the
radiative fountain model (green) to the data. An input power law with
photon index of $\Gamma=2$ is seen through varying layers of column
density (green curves). The data of ESO 103-G035 mostly match some
of the top model curves, but the $8\mathrange10\mathrm{keV}$ data
is underpredicted. This could be accommodated by additional soft energy
emission components. The situation is more discrepant when considering
the Circinus Galaxy. While the data show a powerlaw-like increase
from 8 to 20~keV, the models present a wide hump from 8 to $20\,\mathrm{keV}$.
The models overpredict the data at $10\mathrange25\,\mathrm{keV}$
even at the highest column densities (lowest curve). A better result
may be possible, if an additional reflector was included (see Figure~\ref{fig:modelspectra-wada}).
The data of NGC~424 mimic those of the Circinus Galaxy, but have
larger uncertainties.

The bottom panel of Figure~\ref{fig:data} compares spectra from
the warped disk model (orange) to the data. Here, the data of ESO
103-G035 are directly reproduced by the model, in terms of the $3\mathrange6\mathrm{keV}$
slope, the $8\mathrange20\mathrm{keV}$ shape and the Compton hump
strength. The same is true for NGC~424. For the Circinus Galaxy,
the data can be matched well from $10\,\mathrm{keV}$ upwards. At
energies below $10\,\mathrm{keV}$, an additional contribution appears
to be needed. This is expected. In \citealp{Buchner2019a} and many
other works, an additive powerlaw component has been employed for
this purpose. The photon index $\Gamma$ has not been tuned, which
could lead to even better agreement. The projection of the warped
disk model onto the data is encouraging. To test this more thoroughly,
the multiple spectral components and their model parameters have to
be fitted jointly.

\subsection{Test case: Spectral fit to Circinus\label{sec:circinusspec}}

\begin{figure}
\begin{centering}
\includegraphics[width=1\columnwidth]{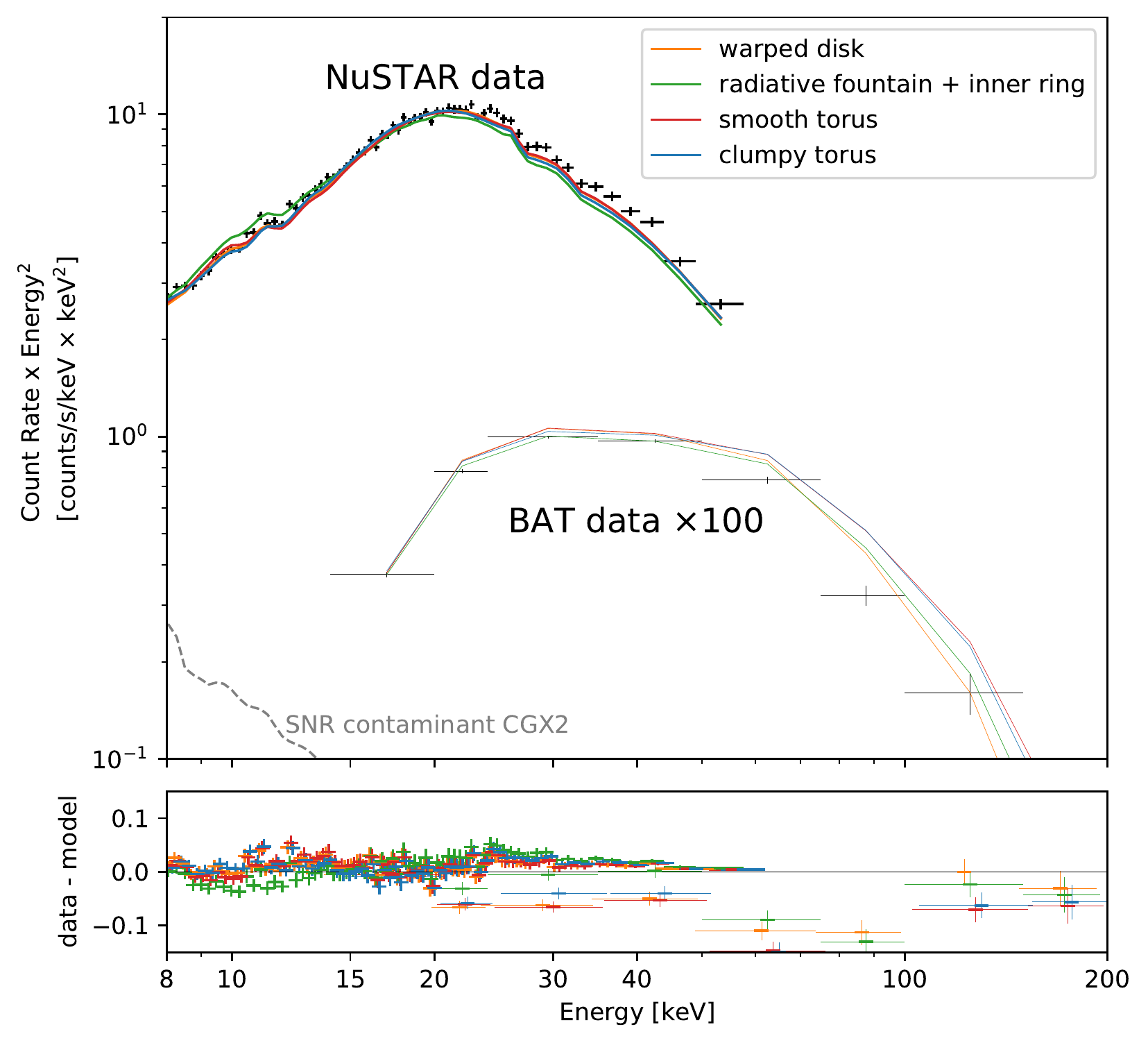}
\par\end{centering}
\caption{\label{fig:circinusspec}Circinus hard X-ray spectrum with fitted
models. The \emph{NuSTAR} $8-75\mathrm{\,keV}$ (here heavily rebinned
just for visualisation) and BAT $15-150\mathrm{\,keV}$spectra were
fitted with four models. The warped disk model yields the lowest $\chi\text{\texttwosuperior}$.}
\end{figure}

\begin{table*}
\caption{\label{tab:Best-fit-parameters}Best-fit models parameters for the
Circinus Galaxy.}

\centering{}%
\begin{tabular}{lccccc}
Model Geometry & $\NH$ & $\Gamma$ & $f_{\mathrm{scat}}$ & $\theta_{\mathrm{incl}}$ & $\chi^{2}$/dof\tabularnewline
 & $10^{24}\mathrm{cm}^{-2}$ &  & $\%$ & $\text{°}$ & \tabularnewline
\hline 
\hline 
\citep[various, in][]{Arevalo2014} & $3\mathrange20$ & $1.6\mathrange2.5$ &  & $\sim75$ & -\tabularnewline
Smooth torus & \nhmytorus & \gammamytorus & \scatmytorus & \inclmytorus & \chidofmytorus\tabularnewline
Clumpy torus & \nhctorus & \gammactorus & \scatctorus & \inclctorus & \chidofctorus\tabularnewline
Warped disk & \nhwarp & \gammawarp & \scatwarp & \inclwarp & \chidofwarp\tabularnewline
Radiative fountain + inner ring & \nhwadaring & \gammawadaring & \scatwadaring & \inclwadaring & \chidofwadaring\tabularnewline
\hline 
\end{tabular}
\end{table*}

Finally, a full spectral fit is performed to the data from the Circinus
Galaxy. This object is most suitable, because the \textbf{hydrodynamic
simulations} of the radiative fountain model have been tailored to
its parameters\textbf{ (accretion rate and black hole mass). Appendix~\ref{sec:Spectral-fits-others}
makes comparisons to the other two galaxies considered in this work.
The observed spectrum from }\textbf{\emph{NuSTAR}}\textbf{ and }\textbf{\emph{Swift}}\textbf{/BAT,
presented in Figure~\ref{fig:circinusspec}, also exhibits high photon
count statistics. For clear presentation, the }\textbf{\emph{NuSTAR}}\textbf{
spectra were summed and rebinned to 1000 counts per bin, but for fitting
we use a spectrum binned to 20 counts per bin. }

Four spectral models of the obscurer geometry are considered. From
this work, the warped disk model and the radiative fountain model
with the inner ring modification are included. The smooth torus geometry
of \citet{MurphyYaqoobMyTorus2009} and the clumpy torus of \citet{Liu2014}
are also considered. In addition to a powerlaw source processed by
an obscurer model, two further components are included. A secondary,
unobscured powerlaw is included with the same photon index and a normalisation
$f_{\mathrm{scat}}$ of up to $10\%$ of the primary powerlaw normalisation.
The CGX2 supernova remnant contamination has been included as a frozen
\texttt{mekal} component with values taken from \citet{Arevalo2014}.
The spectrum was fitted with all parameters controlling the spectral
shape ($\NH$, $\Gamma$, $f_{\mathrm{scat}}$) free to vary. The
normalisation in each data set (FPMA, FPMB, BAT) were left free to
vary independently, to account for cross-calibration factors.

We report the most important spectral parameters in Table~\ref{tab:Best-fit-parameters}.
The fit $\chi^{2}$ statistic is best in the warped disk model ($\chi^{2}/\mathrm{dof}=\chidofwarp$).
The radiative fountain model is a good fit when an inner ring is inserted
($\chi^{2}/\mathrm{dof}=\chidofwadaring$), but not without ($\chi^{2}/\mathrm{dof}=\chidofwada$).
In both models the photon index is $\Gamma=2$, close to typical values
in the population ($\Gamma=1.95\pm0.15$, \citealt{Nandra1994,Ricci2017a}).
The clumpy torus ($\chi^{2}/\mathrm{dof}=\chidofctorus$) and the
smooth torus ($\chi^{2}/\mathrm{dof}=\chidofmytorus$) do not fit
the data as well. The \emph{NuSTAR} data force these models to high,
atypical photon indices ($\Gamma>2.4$), which are disfavored by the
\emph{Swift}/BAT data ($\Gamma=2.09\pm0.02$ in \citealp{Oh2018}).
All models indicate very high column densities ($\NH\geq10^{25}\mathrm{cm}^{-2}$).
For reference, we also list the parameters found in the recent detailed
analysis of \citet{Arevalo2014}, which considered a multitude of
models, including the smooth torus model.

\section{Discussion \& Conclusion}

\label{sec:Discussion}

\begin{figure}
\begin{centering}
\includegraphics[width=1\columnwidth]{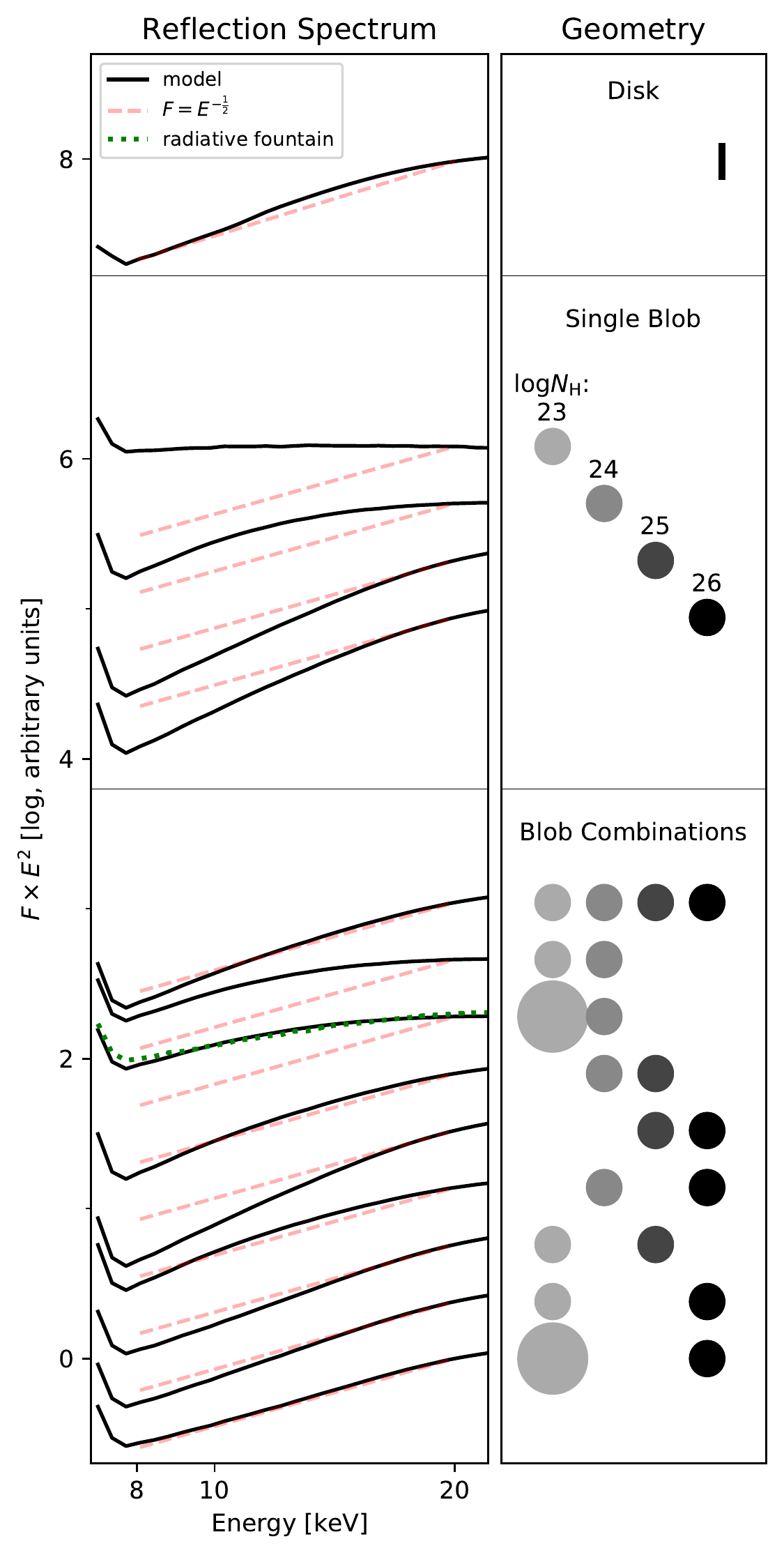}
\par\end{centering}
\caption{\label{fig:Compton-hump-shapes}Reflection spectrum (left, black)
for different geometries (right). Black curves are normalised at $20\,\mathrm{keV}$,
with offsets. The red dashed lines illustrate the observed power law.
The top example shows reflection off a semi-infinite disk. The middle
four examples show a single blob with column densities $\log\NH=23,\,24,\,25$
and $26$. In low-density blobs, the reflection extends to low energies,
while for the densest blobs the spectrum rises rapidly (powerlaw photon
index $\approx0$). The bottom examples show mixtures of the reflection
spectra, assuming equal covering factors in all but the last. Mixtures
have a wider Compton hump with a smooth bend. The exceptions are the
bottom three examples, which lack the $\log\NH=24$ blob. Their powerlaw-like
spectrum approximately follows $F\propto E^{-\frac{1}{2}}$ (red dashed
line). The dotted green line shows the radiative fountain model.}
\end{figure}

In this work we have extended the open source \XARS{} Monte Carlo
spectra simulation software to irradiate arbitrary density grids.
We explore physically motivated geometries, including Wada's radiative
fountain model and warped disks. This allows for the first time investigation
of gas geometries predicted by hydro-dynamic simulations.

\subsection{Geometry information in X-ray spectra}

Different geometries of AGN obscurers predict very different X-ray
spectra. In particular, powerful indicators of the obscurer geometry
in heavily obscured AGN are illustrated by the top left panel of Figure~\ref{fig:modelspectra}.
These include the continuum slope just below the Fe~K emission feature
($2-6\,\mathrm{keV}$), the bend of the continuum between the Fe~K
edge and the Compton hump ($8\mathrange20\,\mathrm{keV}$). Below
we discuss how different geometries produce different shapes.

Spherical, completely covering obscurers predict virtually no escape
of continuum photons below $10\,\mathrm{keV}$. Such geometries have
been suggested for merging galaxies, but are difficult to test because
of the low-energy contamination by stellar processes in these often
star-bursting galaxies \citep[e.g., Arp 220, see][]{Teng2015}.

The reflection processes are more elaborate in other geometries. Warped
disks, sight-lines through the disk heavily suppress the continuum,
while an elevated far side can provide an unabsorbed reflection spectrum.
This behaviour resembles smooth torus geometries \citep{MurphyYaqoobMyTorus2009},
where the far side also provides substantial reflection.

The warped disk structure requires much less mass than a smooth torus,
and is dynamically stable. The extent of the warp is related to the
covering factor and influences the strength of the Compton hump emission
(see Figure~\ref{fig:modelspectra-warp}). 

The radiative fountain model is more complex. A mixture of constant-density
spheres can help understand the emerging spectral shape. The reflection
off a single, spherical blob in the $8\mathrange20~\mathrm{keV}$
band is shown in Figure~\ref{fig:Compton-hump-shapes} (middle four
cases, from the Appendix of \citealp{Buchner2019a}). High-density
blobs show a steeply rising spectrum, while low-density blobs also
let low-energy photons escape. Combining low, intermediate and high-density
blobs is shown in the bottom of Figure~\ref{fig:Compton-hump-shapes}.
The mixture of these reflection spectra leads to a wide, smooth hump.
Such a spectrum emerges from the radiative fountain model (dotted
green curve). In that model, photons interact with a wide range of
mostly Compton-thin densities as they travel towards infinity. The
truncation towards soft energies in Figure~\ref{fig:modelspectra}
can be explained by photo-electric absorption by a Compton-thin gas
envelope. 

\subsection{The obscuring structure of nearby galaxies}

\begin{figure}
\begin{centering}
\includegraphics[width=1\columnwidth]{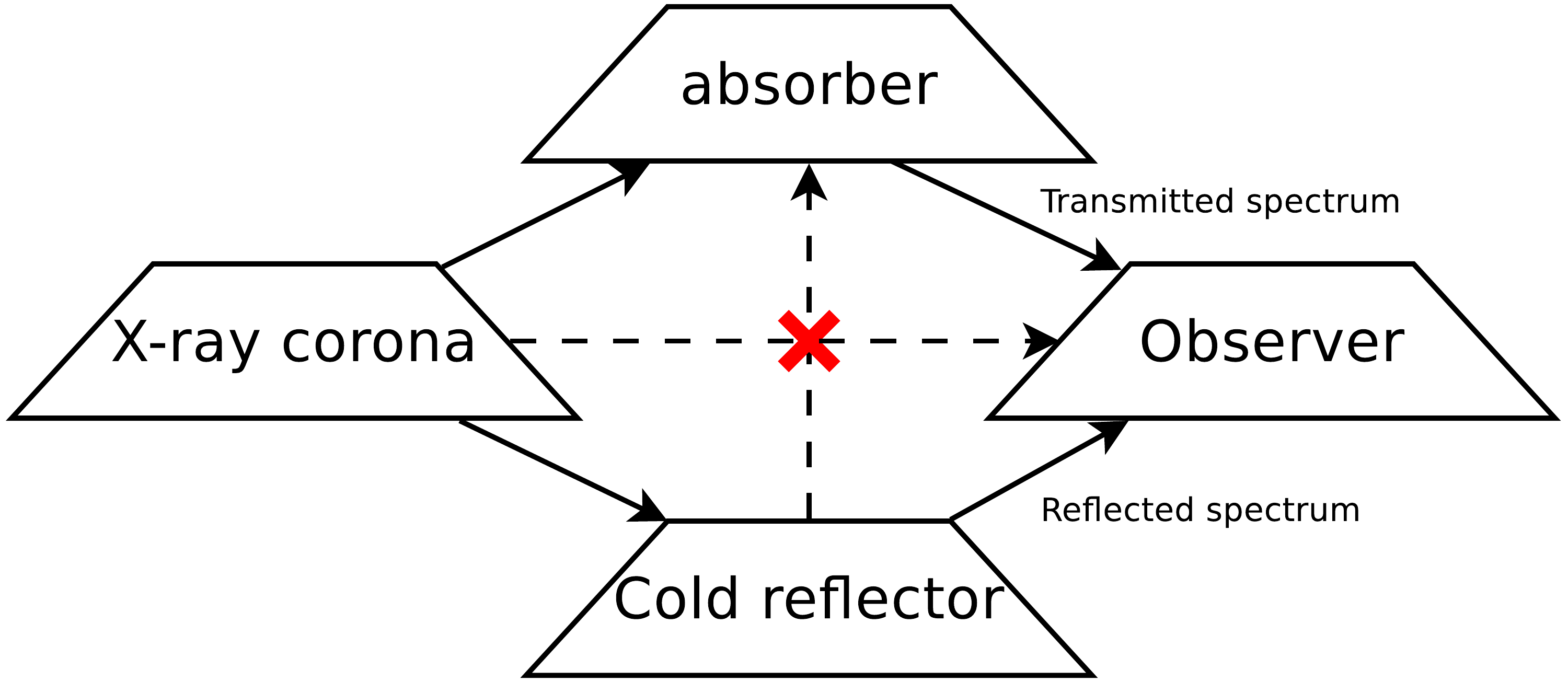}
\par\end{centering}
\caption{\label{fig:losrel}Inferred unobscured line-of-sights in Compton-thick
AGN. Arrows indicate unobscured lines of sight between components.
Importantly, the reflection is not reprocessed by the absorber, and
the corona is not directly observed (red cross). The reflector is
illuminated by the corona directly, not the absorbed emission. The
warm mirror is not shown, but would have similar relations as the
cold reflector. The host and Milky Way galaxy absorbing screens are
also not shown.}
\end{figure}

High-quality high-energy spectra from nearby galaxies enable differentiating
obscurer geometries. Indeed, the warped disk model was selected as
the best fit to the Circinus Galaxy. However, a more generic conclusion
can be drawn. The observed spectrum evinces a power law of $E^{-\frac{1}{2}}$
in the $8\mathrange20\,\mathrm{keV}$ energy window (dashed line in
Figure~\ref{fig:data}). A preliminary analysis of more AGN observed
with \emph{NuSTAR} indicates that such spectral shapes occur only
in rapidly accreting AGN with Eddington accretion ratios of $\dot{M}/\dot{M}_{Edd}\gtrsim20\%$.
The examples in Figure~\ref{fig:Compton-hump-shapes} illustrate
that such photon flux shapes are only possible with a bimodal distribution
of blob densities, including heavily Compton-thick blobs and thinner
blobs $\NH\lesssim10^{23}\mathrm{cm}^{-2}$. For the radiative fountain
model, this suggests that if the thick disk was denser, the model
may follow the data better. Another way to achieve this result, illustrated
by the orange curve in Figure~\ref{fig:modelspectra-wada}, is to
insert an inner high-covering obscuring structure. This effectively
prevents X-ray reflection from the thick disk and alters the spectrum
substantially. A very similar conclusion was reached by \citet{Buchner2019a},
who inserted a inner ring into a clumpy obscurer model.

Figure~\ref{fig:losrel} illustrates the generic recipe to produce
the observed shapes. Firstly, the observed spectrum is reflection-dominated,
therefore the sightline between observer and the X-ray corona features
the Compton-thick obscuration. Secondly, Compton-thick reflecting
surfaces with high covering factors are required to have unobscured
sight-lines both to the observer and the X-ray corona. The accretion
disk is unsuitable for this (see Figure~\ref{fig:modelspectra-wada}).
This is because in non-relativistic, flat accretion disks the disk
reflection and powerlaw both pass through the absorber (dashed, crossed
out arrow in Figure~\ref{fig:losrel}). In contrast, warped disks
are one way to provide Compton-thick reflection surfaces that see
the corona and the observer unobscured (bottom arrows in Figure~\ref{fig:losrel}).
The analysed spectra (Figure~\ref{fig:data} and \ref{fig:Compton-hump-shapes})
also place upper limits on reflection from Compton-thin column densities.

Nevertheless, there are substantial caveats to our conclusions. The
nuclear reflection on parsec and sub-parsec scales needs to be separated
by a combination of \emph{NuSTAR }and \emph{Chandra} observations.
In the future, higher angular resolution missions such as \emph{Lynx}
\citep{TheLynxTeam2018} will provide additional separation. However,
also a wider variety of simulations with higher spatial resolution
is desired. Additionally, ionisation cone imaging spectroscopy (in
optical and X-rays) could be used to constrain the coherence and opening
angle of the inner structure \citep[see e.g.,][]{Zhao2020}.

\subsection{Implications and outlook}

The spectral model diversity in the Compton-thick regime has important
implications. Firstly, the model choice affects the selection of Compton-thick
AGN, and the computation of their space density. Many recent surveys
\citep[e.g.,][]{Ueda2014,Buchner2015,Aird2016} rely on the spectral
model of \citet{Brightman2015}. A fixed spectral model is used to
compute the sensitivity to detecting Compton-thick AGN as a function
of redshift and luminosity. \citet{Ricci2016} explicitly demonstrated
that the derived Compton-thick fraction varies with the assumed geometry.
Thus, knowing the geometry is crucially important. Secondly, hardness
ratios and luminosity absorption corrections in the Compton-thick
regime are strongly dependent on the model geometry, in particular
when analysing band counts (e.g., $0.5\mathrange2\,\mathrm{keV}$
and $2\mathrange10\,\mathrm{keV}$) with a fixed spectral model. Recovering
the intrinsic accretion luminosity is more reliable when using multi-component
spectral fits for each object instead \citep[e.g.,][]{Buchner2014}.

Investigation of spectra of nearby AGN suggests evidence of extreme
column densities of $\NH>10^{25}\mathrm{cm}^{-2}$. These are needed
either along the LOS (see Table~\ref{tab:Best-fit-parameters}) or
are present outside the LOS in the models not excluded by the data
(see Figure~\ref{fig:data} and \citealp{Buchner2019a}). Such high
column densities are outside the range of most previous models. This
is in part because the many photon interactions in such dense media
are computationally costly to evaluate. \XARS{} provides an efficient
implementation for such situations to address this, and can illuminate
arbitrarily complex geometries. In part, such high column densities
($\NH>10^{25}\mathrm{cm}^{-2}$) have not been considered because
they seem to be rare and unlikely a priori. However, current X-ray
surveys of AGN may miss these, as they become strongly absorbed even
in the highest energy surveys (e.g.\emph{, Swift}/BAT), and additional
multi-wavelength selection methods may need to be developed to complement
them. However, strong conclusions require the analysis of a representative
and complete sample. Systematic \emph{NuSTAR} follow-up of the \emph{Swift}/BAT
survey is currently the most promising avenue \citet{Balokovic2017},
with careful multi-component and spatially resolved analyses of \emph{NuSTAR},
\emph{XMM-Newton} and \emph{Chandra} data to separate non-AGN contributions
and extended reflection \citep[as in, e.g.,][]{Arevalo2014,Bauer2014}.
Because of the complex parameter degeneracies (between geometry parameters,
the LOS $\NH$, and source luminosity, $\Gamma$ and $E_{\mathrm{cut}}$),
reliable fits require global parameter exploration algorithms such
as nested sampling \citep[as, e.g., in][]{Buchner2014}.

Different models retrieve different photon indices in the Compton-thick
regime. Even when only $>8\,\mathrm{keV}$ data are considered, the
effect size, $\Delta\Gamma\approx0.4$, is substantially larger than
the uncertainties (see Table~\ref{tab:Best-fit-parameters}). Under
the assumption that the accretion proceeds the same at the same Eddington
ratio, the photon index distribution should be the same in Compton-thin
AGN and Compton-thick AGN. This can provide an additional test of
the geometry, even when multiple geometries provide equally good fits
per-source. For example, \citet{Ueda2014} find different photon indices
for unobscured and moderately obscured AGN, which are not seen in
the local Universe when high-energy data are available \citep{Ricci2017a},
and may point to an incomplete modelling assumption. A similar test
can be done also with the energy cut-off parameter, which can yield
low values ($E_{\mathrm{cut}}<50\,\mathrm{keV}$) in incomplete models.

\section{Acknowledgements}

We thank the anonymous referees for constructive suggestions which
improved the manuscript. 

We acknowledge support from the CONICYT-Chile grants Basal-CATA PFB-06/2007
(JB, FEB), ANID grants CATA-Basal AFB-170002 (JB, FEB), FONDECYT Regular
1141218, 1190818 and 1200495 (all FEB), FONDECYT Postdoctorados 3160439
(JB), ``EMBIGGEN'' Anillo ACT1101 (FEB), and the Ministry of Economy,
Development, and Tourism's Millennium Science Initiative through grant
IC120009, awarded to The Millennium Institute of Astrophysics, MAS
(JB, FEB). This research was supported by the DFG cluster of excellence
``Origin and Structure of the Universe''. MiB acknowledges support
from NASA Headquarters under the NASA Earth and Space Science Fellowship
Program, the YCAA Prize Postdoctoral Fellowship, grant NNX14AQ07H,
and support from the Black Hole Initiative at Harvard University,
which is funded in part by the Gordon and Betty Moore Foundation (grant
GBMF8273) and in part by the John Templeton Foundation. KW was supported
by JSPS KAKENHI Grant Number 16H03959.

This research has made use of the NASA/IPAC Extragalactic Database
(NED), which is operated by the Jet Propulsion Laboratory, California
Institute of Technology, under contract with the National Aeronautics
and Space Administration. This research has made use of SIMBAD \citep{Wenger2000}
and its excellent query API, operated by the Strasbourg astronomical
Data Center (CDS).

\appendix

\bibliographystyle{aa}
\bibliography{agn}

\section{3D Discrete Digital Analyser\label{sec:3DDDA}}

\begin{algorithm}
\caption{\label{fig:DDA}A simplified 2D version of our grid traversal algorithm.}

\begin{lstlisting}[language=C,basicstyle={\scriptsize},breaklines=true,tabsize=4]
function DDA(
	// starting position
	double x0, double y0, 
	// direction
	double dx, double dy,
	// column density
	double NH
) {

// initialisation at current cell
double dt_dx = 1. / dx;
double dt_dy = 1. / dy;

int x, y = x0, y0;
int x_inc, y_inc;
double t_next_x, t_next_y;
int sign_x = 1, sign_y = 1;

if (dx > 0) {
	x_inc = 1;
	t_next_x = (floor(x0) + 1 - x0) / dx;
} else {
	x_inc = -1;
	t_next_x = (x0 - floor(x0)) / dx;
	sign_x = -1;
}
if (dy > 0) {
	y_inc = 1;
	t_next_y = (floor(y0) + 1 - y0) / dy;
} else {
	y_inc = -1;
	t_next_y = (y0 - floor(y0)) / dy;
	sign_y = -1;
}
loop {
    double last_t = t; int last_x = x, last_y = y;
	if (t_next_y*sign_y <= t_next_x*sign_x) {
		// go in y
		y += y_inc;
		t = t_next_y;
		t_next_y += dt_dy;
	} else {
		// go in x
		x += x_inc;
		t = t_next_x;
		t_next_x += dt_dx;
	}
	// density of this grid cell:
	double rho = rhoarr[last_x, last_y];
	if ((fabs(t) - last_t) * rho > NH) {
		// terminate within this cell.
        // return distance travelled
		return last_t + NH / rho;
	}
	// subtract full cell
	NH -= (fabs(t) - last_t) * rho;
} // end of loop
} // end of function
\end{lstlisting}

\end{algorithm}

\begin{figure}
\begin{centering}
\includegraphics[width=1\columnwidth]{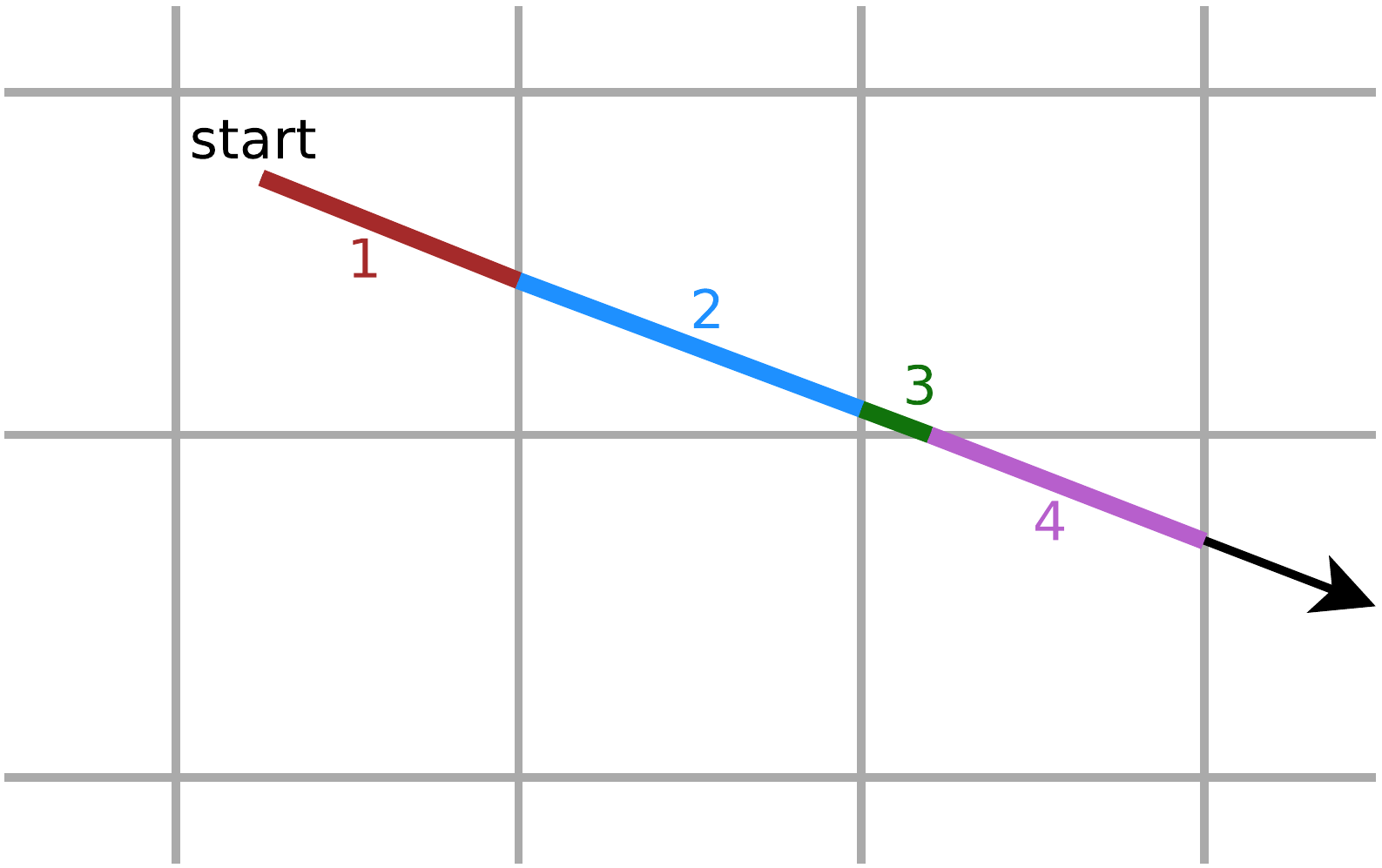}
\par\end{centering}
\caption{\label{fig:DDA-Illustration}Illustration of photon propagation in
a grid. The Discrete Digital Analyser algorithm finds the sequence
of points where the line crosses grid boundaries. This allows segmentation
of the line by grid cell, illustrated by the numbers and colours.}

\end{figure}

We briefly describe how density grids were illuminated and processed
into X-ray spectra. In \XARS{} \citep[see][]{Buchner2019a}, the
user needs to specify two functions, one describing the geometry,
and the second specifying how escaping photons are gridded (e.g. in
viewing angle). For the latter we chose gridding in viewing angle
and LOS column density to the corona \citep[see][]{Buchner2019a},
as these are the primary aspects affecting the output spectrum. The
custom geometry function receives a photon position, direction and
$\NH$ distance (in optical depth units) to travel until the photons
stops (for interaction with the medium).

In the case of our density grids, the density within a grid cell is
assumed to be constant. The problem then transforms to measuring the
distances traversed in each grid cell and multiplying with the grid
density to obtain the grid cells $\NH$. We illustrate the segmentation
of the path in Figure~\ref{fig:DDA-Illustration}. This is subtracted
from the target $\NH$ distance until it reaches zero. Measuring the
distances traversed in each grid cell is solved with a 3D Discrete
Digital Analyser, an algorithm developed for drawing lines through
pixels in computer graphics \citep{4056861}. We present a two-dimensional
simplified version in Figure~\ref{fig:DDA}; the extension to three
dimensions is straight-forward. The algorithm computes the step size
between grid lines given the current direction, and then steps through
the grid cells in the right order (either in x or y). The outlined
algorithm returns the coordinate space distance traversed until the
photon has travelled through a column density of $\NH$. 

\section{Spectral fits of NGC~3393 and NGC~424}

\label{sec:Spectral-fits-others}For completeness, we also present
spectral fits for the two other galaxies, NGC~3393 and NGC~424.
The setup is virtually identical to that described in section §\ref{sec:circinusspec}.
The only difference in the modelling is that no SNR contaminant is
included. 

\begin{figure}
\includegraphics[width=1\linewidth]{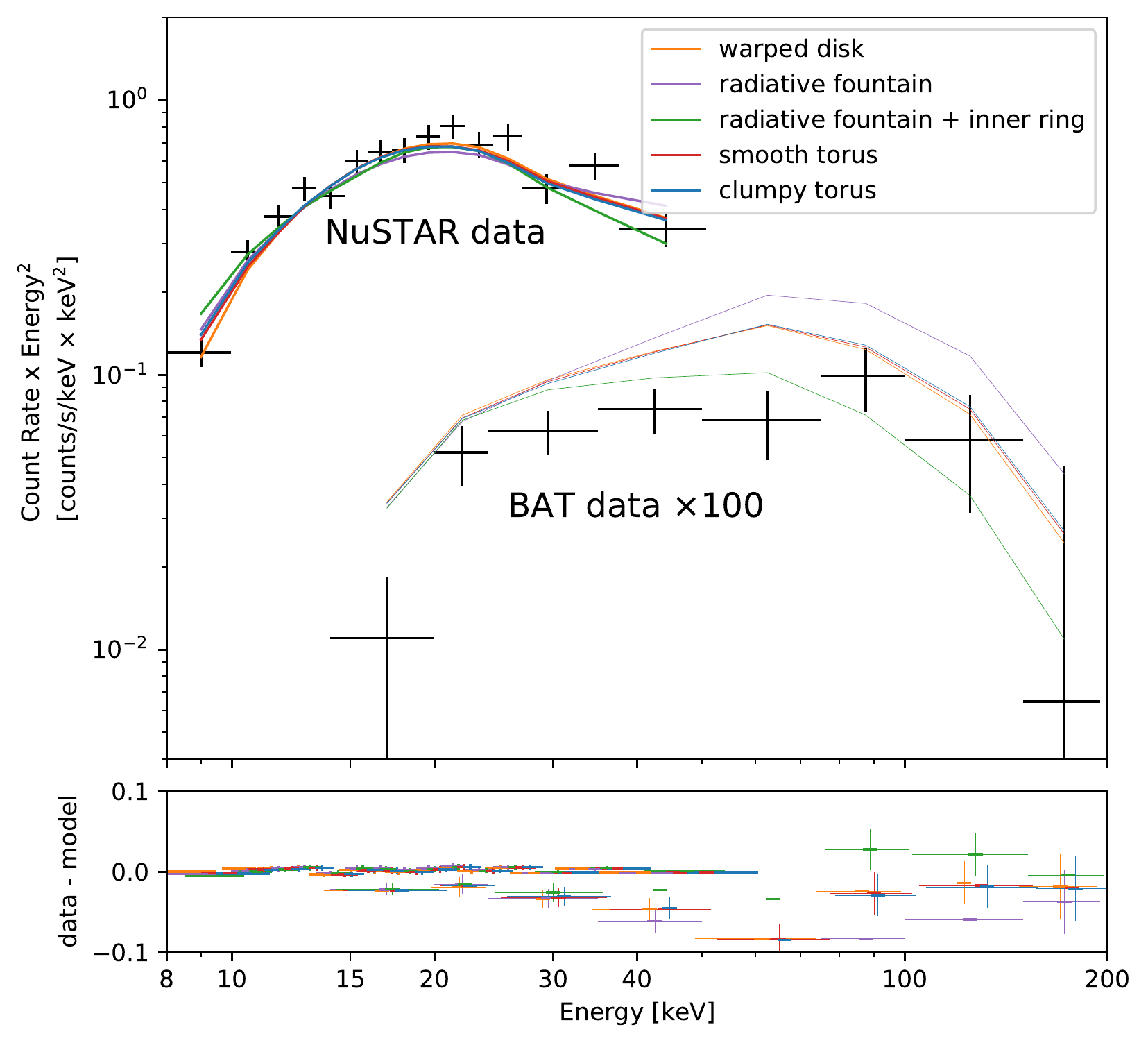}

\caption{\label{fig:NGC3393fit}Same as Figure~\ref{fig:circinusspec}, but
for NGC~3393.}

\end{figure}

Figure~\ref{fig:NGC3393fit} presents the spectral fit of five models.
All models give good fit qualities ($\chi^{2}/\mathrm{dof}<80/77$).
However, the best fit photon indices of the clumpy model and the radiative
fountain model are low ($1.69\text{\ensuremath{\pm}}0.11$ and $1.53\pm0.16$,
respectively). For the other models, including the radiative fountain
model with inner ring, the best fit photon index has values more typical
of the population (the $1.8-2.0$ interval overlaps the $1\sigma$
confidence intervals). All model fits produce best-fit column densities
in the Compton-thick regime.

\begin{figure}
\includegraphics[width=1\linewidth]{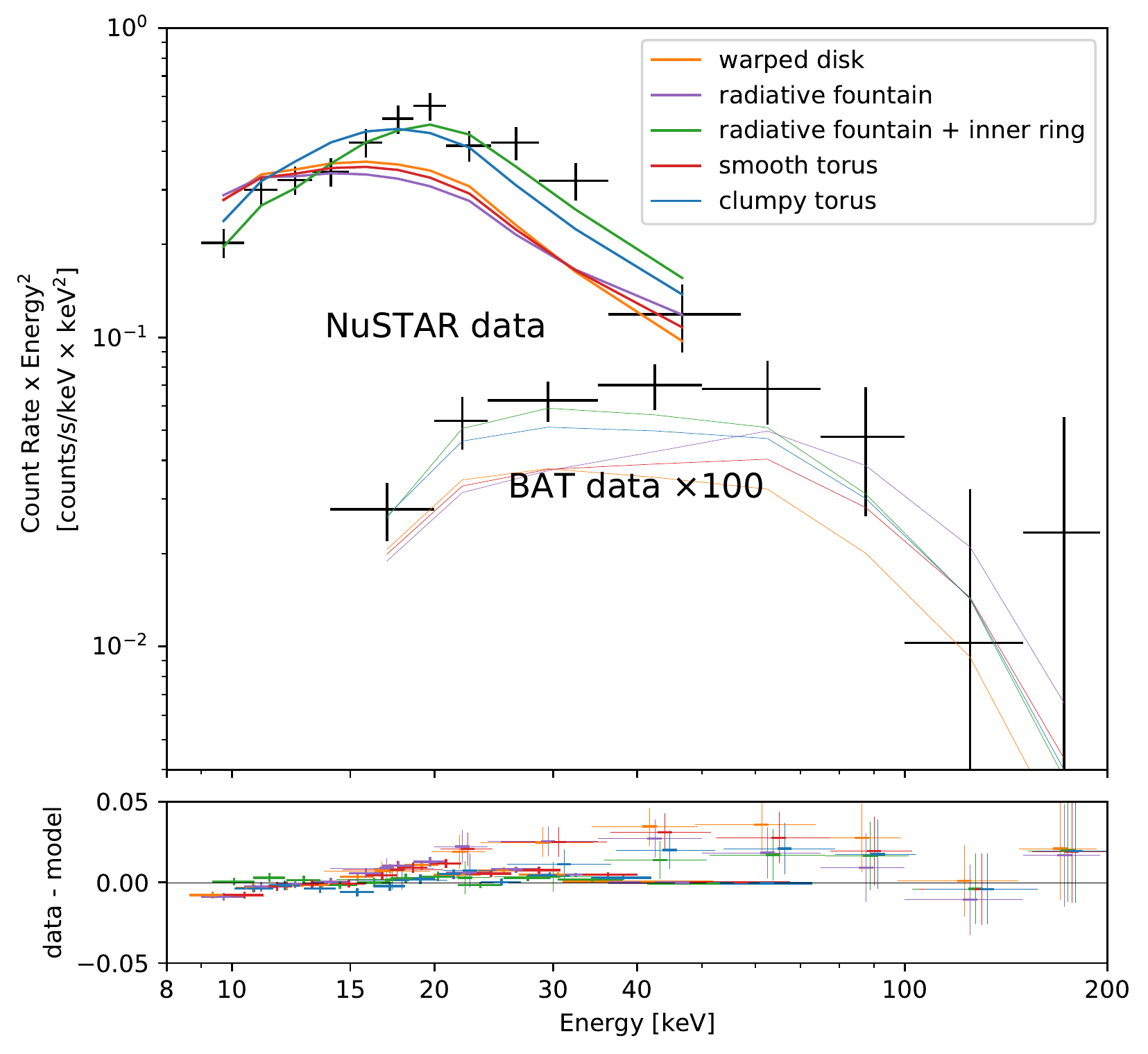}

\caption{\label{fig:NGC424fit}Same as Figure~\ref{fig:circinusspec}, but
for NGC~424.}
\end{figure}

Figure~\ref{fig:NGC424fit} presents the equivalent spectral fit
for NGC~424. All models give good fit qualities ($\chi^{2}/\mathrm{dof}\lesssim115/128$).
However, the best fit photon indices of the clumpy torus model and
the smooth torus model are high ($2.46\text{\ensuremath{\pm}}0.27$
and $2.45\pm0.26$, respectively). For the other models, the best
fit photon index hav values typical of the population ($2.0$ is within
the $1\sigma$ confidence intervals). All model fits produce best-fit
column densities in the Compton-thick regime.

The spectral fits confirm the point of Figure~\ref{fig:data}: The
spectra can be explained by an AGN with a typical photon index $\Gamma=2$
under the warped disk model, but not under the radiative fountain
model, unless a inner ring is added.

\end{document}